\documentclass[11pt,a4paper]{article}
\pdfoutput=1
\usepackage{jheppub}
\usepackage{rotating}
\usepackage{array}
\usepackage{amsmath}
\usepackage[normalem]{ulem}
\usepackage{slashed}
\usepackage{booktabs}
\usepackage[pdftex,table,dvipsnames]{xcolor}
\usepackage{units}
\usepackage{multirow}
\usepackage{url}
\usepackage{parskip}
\usepackage[all]{nowidow}
\usepackage{placeins}

\usepackage{longtable}

\usepackage{xspace}
\usepackage{subfig}
\usepackage{color}

\usepackage{dsfont}
\usepackage{soul}
\usepackage{hyperref}

\usepackage{physics} % For some physics symbols and better parenthesis
\usepackage{bm} % For bold math symbols

\allowdisplaybreaks

\newcommand{\bc}{\bm{c}}

\newcommand{\mL}{\mathcal{L}}
\newcommand{\SMEFT}{\mathrm{SMEFT}}
\newcommand{\hc}{\mathrm{h.c.}}

\def \cgg {\tilde{c}_{GG}}
\def \ctt {c_{tt}}
\newcommand{\mM}{\mathcal{M}}
\newcommand\figref{Figure~\ref}

\preprint{Nikhef 2023-021}

\title{Precise tests of the axion coupling to tops}

\author[1,2]{A. V. Phan}
\author[1,2]{S. Westhoff}

\affiliation[1]{Institute for Mathematics, Astrophysics and Particle Physics, Radboud University, 6500 GL Nijmegen, The Netherlands}
\affiliation[2]{Nikhef, Science Park 105, 1098 XG Amsterdam, The Netherlands}

\emailAdd{anhvu.phan@ru.nl}
\emailAdd{susanne.westhoff@ru.nl}

\abstract{We present an in-depth analysis of axions and axion-like particles in top-pair production at the LHC. Our main goal is to probe the axion coupling to top quarks at high energies. To this end, we calculate the top-antitop cross section and differential distributions including ALP effects up to one-loop level. By comparing these predictions with LHC precision measurements, we constrain the top coupling of axion-like particles with masses below the top-antitop threshold. Our results apply to all UV completions of the ALP effective theory with dominant couplings to top quarks, in particular to DFSZ-like axion models.}

\begin{document}

\maketitle

\flushbottom

%%%%%%%%%%%%%%%%%%%%%%%%%%%%%%%%%%%%%%%%%%%%%%%%%%%%%%%%%%%%%%%
\section{Introduction}
\label{SEC:introduction}
Axions and axion-like particles (ALPs) are omnipresent in extensions of the Standard Model (SM). The strong CP problem and its axion solution is certainly the most famous example of such an extension~\cite{Peccei:1977hh,Peccei:1977ur}. But axion-like particles, that is, pseudo-scalars with shift-symmetric couplings, are also predicted in many models that do not address the strong CP problem. Whenever a theory involves a spontaneously broken global symmetry, the corresponding pseudo-Goldstone boson is an axion-like particle.

The shift symmetry of the ALP Lagrangian implies that ALP couplings to fermions must be proportional to the fermion mass. This common feature of all ALPs distinguishes them from more general pseudo-scalars. ALP couplings to gauge bosons, as well as their relative strength compared to fermion couplings, are very model-dependent~\cite{DiLuzio:2020wdo,Arias-Aragon:2022iwl}. For example, in DFSZ-like axion models~\cite{Dine:1981rt,Zhitnitsky:1980tq} couplings to gauge bosons are loop-induced by the fermion couplings, while in KSVZ models~\cite{Kim:1979if,Shifman:1979if} fermion couplings are generally suppressed.

The flavor hierarchy of the ALP-fermion couplings gives the top quark a special role in the phenomenology of ALPs. Since top quarks interact with all bosons of the Standard Model, the strong ALP-top coupling affects the energy evolution of all other ALP couplings~\cite{Choi:2017gpf,MartinCamalich:2020dfe,Chala:2020wvs,Bauer:2020jbp}. Probing the ALP-top coupling at high energies informs us about possible effects at lower energies. In particular, in models where the ALP-gauge couplings are induced by fermion loops, probing the ALP-top coupling implies probing the ALP-gauge couplings.

In this work, we perform a detailed analysis of ALP effects in top-antitop production at the LHC. Precisely measured top observables allow us to probe the ALP-top coupling directly with resonant top quarks. Moreover, ALP effects in top-antitop production are largely independent of the ALP mass, as long as it lies below the top-antitop threshold. This makes our analysis applicable to a broad range of ALP models. Above threshold, resonance searches in top-antitop production offer a high sensitivity, see e.g.~\cite{Barger:2006hm,Frederix:2007gi,Bai:2008sk,Bonilla:2021ufe}. Searches for top-antitop production plus missing energy~\cite{Esser:2023fdo} or a pseudo-scalar resonance~\cite{Rygaard:2023vyo} can also efficiently probe the ALP-top coupling, but require that the ALP is light enough to decay invisibly or displaced. For light ALPs, observables with virtual top quarks like rare meson decays~\cite{Batell:2009jf,Dolan:2014ska,Bauer:2021mvw,Ferber:2022rsf} or Higgs decays~\cite{Bauer:2017ris} have a high sensitivity to the ALP-top coupling.

Partial effects of ALPs in top-antitop production have been previously explored~\cite{Galda:2021hbr,Bonilla:2021ufe,Esser:2023fdo,Bruggisser:2023npd}. Our analysis goes beyond these studies in two ways: First, we perform a full calculation of ALP effects in top-antitop production at tree level and one-loop level. Given the precision in top observables, this is necessary to obtain robust predictions of the ALP effects. Second, we investigate the impact of the ALP-gluon coupling, which enters the amplitudes at tree level. We show how the sensitivity of top observables to the ALP-top coupling depends on the magnitude and sign of the ALP-gluon coupling. This allows us to probe a large class of ALPs and in particular DFSZ-like axion models, where the ALP-gluon is induced by the ALP-top coupling.

This article is organized as follows. In Sec.~\ref{SEC:alps}, we recapitulate the effective theory of ALP interactions and set the stage for our analysis. In Sec.~\ref{SEC:effects}, we describe our calculation of the various ALP effects in top-antitop production in detail and present numerical predictions of differential distributions. Analytic expressions are given in App.~\ref{app:loop-results}. In Sec.~\ref{SEC:results}, we compare these predictions to LHC data and derive bounds on the ALP coupling to top quarks and gluons. In Sec.~\ref{SEC:comparison}, we compare our results to other observables that probe the ALP-top coupling. We conclude in Sec.~\ref{SEC:conclusions}.

Shortly before finishing this work, Ref.~\cite{Blasi:2023hvb} appeared, which includes an analysis of ALP effects in top-antitop production beyond tree level, as well as other top processes. Since the setup and focus are different from ours, we do not attempt to perform a detailed comparison, but refer the interested readers to the article.

%%%%%%%%%%%%%%%%%%%%%%%%%%%%%%%%%%%%%%%%%%%%%%%%%%%%%%%%%%%%%%%
\section{ALP effective theory}
\label{SEC:alps}
ALPs, by definition, are pseudo-scalars whose interactions preserve
a shift symmetry $a \to a + c$, where $a$ is the ALP field and $c$ is a constant. The leading shift-symmetric interactions are of mass dimension 5, described by an effective field theory (EFT) with a cutoff scale $\Lambda$~\cite{Georgi:1986df}. At energies $\mu$ below the cutoff scale, but above the weak scale, the effective Lagrangian for ALP interactions relevant to top-antitop production reads
\begin{align}\label{eq:lagrangianI}
    \mathcal{L}_{I}(\mu) & = \frac{1}{2}\,\partial_\mu a\,\partial^\mu a - \frac{m_a^2}{2}\,a^2 - \frac{\partial^\mu a}{f_a} \sum_{F=Q,U,D} \bar F\, \bc_F \gamma_\mu F + c_{GG}\,\frac{a}{f_a}\,\frac{\alpha_s}{4\pi}\,G_{\mu\nu}^A \widetilde{G}^{A,\mu\nu},
\end{align}
with the gluon field strength tensor $G_{\mu\nu}^A$ and its dual $\widetilde{G}^{A,\mu\nu} = \frac{1}{2}\epsilon^{\mu\nu\rho\sigma}G_{\rho\sigma}^A$ with $SU(3)_C$ gauge indices $A = 1,\dots 8$. All parameters in this Lagrangian are defined at the scale $\mu$. We have adopted the notation of Ref.~\cite{Bauer:2020jbp}, which sets $\Lambda = 4\pi f_a$. Throughout this work, we set $f_a = 1\,$TeV.  For UV completions with $\Lambda = f_a$, our results can be interpreted by simply replacing $f_a \to 4\pi f_a$. The ALP mass $m_a$ breaks the shift symmetry. We consider it as a free parameter and do not speculate about its origin.

The shift symmetry of the ALP-fermion couplings is explicit through the partial derivative $\partial^\mu a$. The symmetry imposes a particular structure on the $3\times 3$ coupling matrix in flavor space, $\bc_F$, where $F=Q,U,D$ denote electroweak doublet, up-type and down-type gauge singlet quarks. In strong-interaction processes like top-antitop production, the ALP-fermion vector current is conserved and only axial-vector couplings are observable. We define the flavor-diagonal axial-vector couplings to up-type and down-type quarks as ($i = {1,2,3}$)
\begin{align}
c_{ii}^{(u)} & = (\bc_U)_{ii} - (\bc_Q)_{ii},\qquad c_{33}^{(u)} \equiv \ctt,\\\nonumber
c_{ii}^{(d)} & = (\bc_D)_{ii} - (\bc_Q)_{ii},\qquad c_{33}^{(d)} \equiv c_{bb}.
\end{align}

While the Lagrangian~\eqref{eq:lagrangianI} makes the shift symmetry explicit, it can be (and in our case will be) convenient to work in a different basis with Lagrangian
\begin{align}\label{eq:lagrangianII}
    \mathcal{L}_{I\!I}(\mu) & = \frac{1}{2}\,\partial_\mu a\,\partial^\mu a - \frac{m_a^2}{2}\,a^2 - \frac{a}{f_a}\left(\overline{Q} \widetilde{H}\, \widetilde{\bf{Y}}_U\,U +\overline{Q}H\, \widetilde{\bf{Y}}_D\,D + h.c. \right)\\\nonumber
    & \quad + \cgg \,\frac{a}{f_a}\,\frac{\alpha_s}{4\pi}\,G_{\mu\nu}^A \widetilde{G}^{A,\mu\nu}\,,
\end{align}
where $\widetilde{H} = i\sigma_2 H^\ast$ with the Higgs doublet $H$. In what follows, we will refer to~\eqref{eq:lagrangianI} and \eqref{eq:lagrangianII} as Basis I and Basis II, respectively.

In Basis II, the ALP-fermion couplings are given by
\begin{align}
\widetilde{\bf{Y}}_U & = i\left({\bf Y}_U \bc_U - \bc_Q {\bf Y}_U\right),\qquad
\widetilde{\bf{Y}}_D = i\left({\bf Y}_D \bc_D - \bc_Q {\bf Y}_D\right),
\end{align}
with the up-type and down-type Yukawa matrices ${\bf Y}_U$ and ${\bf Y}_D$. In the basis where the interaction and mass eigenstates of up-type quarks coindice, ${\bf Y}_U = \hat{{\bf Y}}_U$ is diagonal and ${\bf Y}_D = {\bf V}\hat{{\bf Y}}_D$, where ${\bf V}$ is the CKM matrix. Neglecting the light-quark Yukawa couplings, the ALP-quark and ALP-gluon couplings in~\eqref{eq:lagrangianI} and~\eqref{eq:lagrangianII} are then related by ($i,j,k = 1,2,3$)
\begin{align}\label{eq:translation}
(\widetilde{\bf{Y}}_U)_{ij} & = i\,y_t\,\Big((\bc_U)_{ij}\delta_{i3} - (\bc_Q)_{ij}\delta_{j3}\Big)\\\nonumber
(\widetilde{\bf{Y}}_D)_{ij} & = i\,y_b\,\Big(V_{i3}(\bc_D)_{3j} - (\bc_Q)_{ik}V_{k3}\delta_{j3}\Big)\\\nonumber
\tilde{c}_{GG} & = c_{GG} + \frac{1}{2}\sum_i \Big( (\bc_U)_{ii} + (\bc_D)_{ii} - 2(\bc_Q)_{ii} \Big).
\end{align}
These relations apply at all scales above the top mass scale. They can be used to translate between the parameters in Basis I and Basis II.

Assuming that the ALP couples only to top quarks (and not to lighter quarks), at the cutoff scale $\Lambda$ the model parameters in Basis I and II are related as
\begin{align} \label{eq:basis-change}
    (\widetilde{\bf{Y}}_U)_{33} = i\,y_t\,c_{tt},\qquad  \cgg = c_{GG} + \frac{\ctt}{2}.
\end{align}
In top-antitop production, this is a realistic simplification, since ALP couplings to on-shell up and charm quarks are suppressed by the small quark masses. Flavor off-diagonal couplings would leave traces in top observables, but we do not consider them in this work.

The ALP couplings $\ctt(\mu)$ and $\cgg(\mu)$ are scale-dependent quantities. Their evolution with energy is described by a coupled set of renormalization group (RG) equations. At leading order in perturbation theory, the solution of the RG evolution for scales $\mu < \Lambda$ is approximately given by~\cite{Bauer:2020jbp}
\begin{align}\label{eq:running}
\ctt(\mu) & = \left[1 - \frac92 R(\mu,\Lambda)\right] \ctt(\Lambda)\\\nonumber
\cgg(\mu) & = \cgg(\Lambda) - R(\mu,\Lambda)\, \ctt(\Lambda),
\end{align}
with the evolution function
\begin{align}
R(\mu,\Lambda) = \frac{\alpha_t(\mu)}{\alpha_s(\mu)}\left[1-\left(\frac{\alpha_s(\Lambda)}{\alpha_s(\mu)}\right)^{\frac{1}{7}}\right],\qquad \alpha_t(\mu) = \frac{y_t^2(\mu)}{4\pi}.
\end{align}
Setting $\ctt(\Lambda) = 1$ and $\cgg(\Lambda) = 0$ at $\Lambda = 4\pi\,$TeV, we obtain the effective couplings at the top mass scale $\mu = m_t = 172.5\,$GeV as
\begin{align}\label{eq:running-num}
\ctt(m_t) = 0.81,\qquad c_{bb}(m_t) = 0.10,\qquad \cgg(m_t) = -0.04.
\end{align}
The RG evolution of $\ctt$ reduces the ALP-top coupling and induces an ALP-gluon coupling at scales below the cutoff scale. The numerical values are consistent with~\eqref{eq:translation}; RG-induced contributions of first- and second-generation quarks cancel in the sum. In our numerical analysis, we take this effects into account. The effect of $\cgg(\Lambda)$ on the RG evolution of $\ctt$ is suppressed as $\alpha_s^2$ and will be neglected.

The main goal of this work is to derive a robust bound on the ALP-top coupling. In the main part of our analysis, we will focus on the scenario where only top couplings are present at the cutoff scale. In terms of the parameters from Basis I, we define
\begin{align}\label{eq:benchmarkI}
\text{Benchmark I:} \qquad c_{GG}(\Lambda) = 0,\quad \ctt(\Lambda) = \ctt.
\end{align}
Using~\eqref{eq:basis-change}, this corresponds to $\cgg(\Lambda) = \ctt/2$ in Basis II.

Since Basis II is also often used in the literature, we define a second benchmark
\begin{align}\label{eq:benchmarkII}
\text{Benchmark II:} \qquad \cgg(\Lambda) = 0,\quad (\widetilde{\bf{Y}}_U)_{33}(\Lambda) = i\,y_t\,\ctt.
\end{align}
Such a scenario is realized in DFSZ-like axion models~\cite{Dine:1981rt,Zhitnitsky:1980tq}, where the axion-fermion couplings are Yukawa-like and couplings to SM gauge bosons are induced only through fermion loops. Notice that the two benchmarks do not describe the same class of physics models. In what follows, the parameters $\ctt$, $\cgg$ and $c_{GG}$ are always defined at the cutoff scale $\Lambda$, unless mentioned otherwise.

These two benchmarks serve as reference points for our analysis. In Secs.~\ref{sec:gluon-coupling-and-mass} and~\ref{sec:ctt-cgg}, we extend the analysis to arbitrary ALP-gluon couplings. In this way, our results are applicable to general UV completions with top and gluon couplings, interpreted in the 2-dimensional parameter space of $\{\ctt,c_{GG}\}$.

%%%%%%%%%%%%%%%%%%%%%%%%%%%%%%%%%%%%%%%%%%%%%%%%%%%%%%%%%%%%%%%
\section{ALP effects in top-antitop production}
\label{SEC:effects}
One of the best processes to probe the ALP coupling to top quarks is top-antitop production at the LHC. The resonant production of the top-antitop pair allows us to test the ALP-top coupling directly, compared to observables like meson decays or Higgs decays, where the ALP couples to virtual tops. The main kinematic distributions of hadronic top-antitop production have been measured with a precision of a few percent. This makes them sensitive to subtle effects of light new particles like ALPs.

Throughout this section, we work in Basis II from~\eqref{eq:lagrangianII}, which is better suited to calculate ALP effects beyond tree level.

\subsection{ALP contributions at LO and NLO}
\label{sec:lo-nlo}
\begin{figure}[t!]
    \centering
   \includegraphics[width=0.8\textwidth]{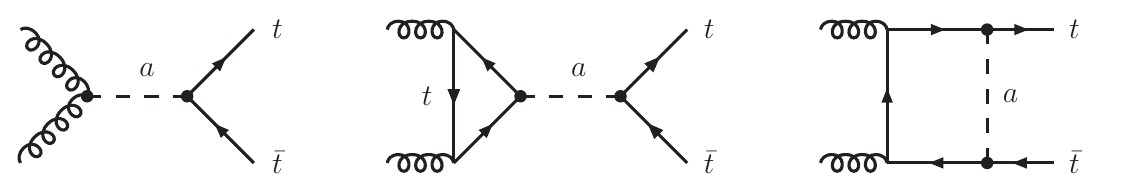}\\[0.4cm]
   \includegraphics[width=0.8\textwidth]{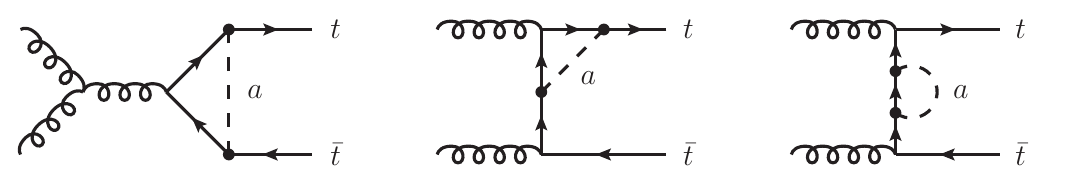}\\[0.4cm]
   \includegraphics[width=0.8\textwidth]{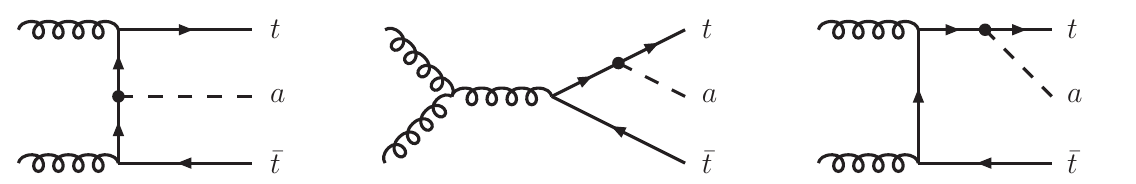}
    \caption{ALP contributions to top-antitop production at LO (upper left diagram) and at NLO from virtual corrections (first and second row) and from real radiation (third row). Only representative Feynman diagrams of each class are shown.}\label{fig:diagrams}
\end{figure}

ALPs leave various effects in top-antitop production. Since ALP couplings to external light quarks are suppressed as $m_q/f_a$, we neglect quark-induced processes and focus on the partonic amplitude $gg\to t\bar{t}$. We include all $\ctt$-induced ALP effects, as well as effects of $\cgg$ at tree level. As we will see, this allows us to test all ALP scenarios with couplings $|\ctt| > \alpha_s|\cgg|/4\pi$ at the cutoff scale.
%the Benchmark~\eqref{eq:benchmarkII} with $\cgg(\Lambda) = 0$ at the cutoff scale.
 In~\figref{fig:diagrams}, we show representative Feynman diagrams for the various ALP contributions to inclusive top-antitop production.

To classify these contributions, it is useful to distinguish between predictions in QCD (``SM'') and predictions in QCD plus effective ALP interactions (``SM+ALP''). At leading order (LO) in perturbation theory of the SM+ALP, ALP contributions to observables are proportional to $\ctt\cgg$. At next-to-leading order (NLO), ALP contributions through virtual corrections and real radiation are proportional to $\ctt^2$. In what follows, we will discuss the ALP contributions at LO and NLO one by one.

%%%%%%%%%%%%%%%%%%%%%%%%%%%%%%%%%%%
%%%%%%%%%%%%%%%%%%%%%%%%%%%%%%%%%%%

%---------------------------------------------------------------
\paragraph{LO contributions}
The amplitude for $gg\to t\bar{t}$ in the SM+ALP is sensitive to the product of couplings $\cgg(\mu)\ctt(\mu)/f_a^2$. Even if the ALP-gluon coupling is absent at the cutoff scale, as in the benchmarks from~\eqref{eq:benchmarkI} and~\eqref{eq:benchmarkII}, it can still be induced by the RG evolution of the ALP-top coupling, see Sec.~\ref{SEC:alps}. At scales $\mu < \Lambda$, the RG-induced ALP-gluon coupling is roughly given by
\begin{align}\label{eq:ll}
    \cgg(\mu) \approx \frac{\alpha_t}{4\pi}\, \ctt(\Lambda)\ln\left(\frac{\Lambda^2}{\mu^2}\right).
\end{align}
In Benchmark II with $\ctt(4\pi\text{TeV}) = 1$, we find $\cgg(m_t) \approx - 0.04$ when resumming the UV-sensitive logarithmic contributions to all orders. In the LO amplitude for $gg\to t\bar{t}$, the RG-induced ALP-gluon coupling leads to an ALP effect of
\begin{align}
    \mathcal{O}\left(\frac{\cgg(\mu) \ctt(\mu)}{f_a^2}\right) \sim \frac{\alpha_t}{4\pi} \frac{\ctt^2(\Lambda)}{f_a^2}\ln\left(\frac{\Lambda^2}{\mu^2}\right).
\end{align}
We will denote LO contributions to $t\bar{t}$ observables by $\sigma^{(t\bar{t},0)}$.

%---------------------------------------------------------------
\paragraph{Virtual NLO contributions}
At NLO in the SM+ALP, ALPs contribute to $gg\to t\bar{t}$ through virtual and real corrections, as shown in \figref{fig:diagrams}. The one-loop diagrams in the first row are UV-finite. The diagrams in the second row (and topologically similar contributions) are UV-divergent and require renormalization.

To absorb the UV divergences in the one-loop $gg\to t\bar{t}$ amplitude, we introduce counter\-terms for the top-antitop-gluon vertex and for the top-quark two-point function,
\begin{center}
\begin{tabular}{ m{4cm} p{7cm} }
    \includegraphics[width=4cm]{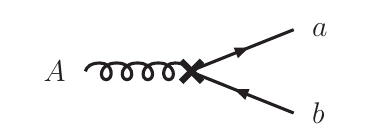} &$i g_s \delta_1 \gamma^\mu T_{ab}^A$ \\
    \includegraphics[width=4cm]{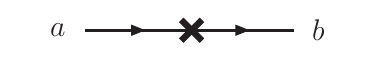} &$i \big(\delta_2 \slashed{p} - (\delta_2 + \delta_m) m_t\big)\delta_{ab}$,\\
\end{tabular}
\end{center}
where $g_s$ is the strong coupling constant, $T_{ab}^A$ is a generator of $SU(3)_C$, and $p$ is the four-momentum of the top quark. The parameters $\delta_m$, $\delta_1$ and $\delta_2$ are the renormalization constants for the top mass, the $t\bar{t}g$ vertex and the top field strength. Due to the Ward-Takahashi identity, $\delta_1 = \delta_2$. The counterterms can thus be determined by imposing two renormalization conditions.

For the top mass renormalization, we choose the on-shell scheme, which is appropriate for amplitudes with on-shell tops. To renormalize the top field strength, we work in the $\overline{\text{MS}}$ scheme. The two renormalization conditions read
\begin{align}
\hat{\Sigma}^{ab}_t(\slashed{p} = m_t) = 0,\qquad \frac{d}{d\slashed{p}}\eval{\hat{\Sigma}^{ab}_t(\slashed{p})}_{\mathrm{div}} = 0,
\end{align}
where $\hat{\Sigma}^{ab}_t(\slashed{p})$ is the renormalized two-point function of the top quark and `div' refers to the divergent part of its partial derivative.

Applying these conditions, the renormalized two-point function is given by
\begin{align}
\hat{\Sigma}^{ab}_t(\slashed{p}) = \Sigma^{ab}_t(\slashed{p}) + \delta_{ab}\big(\delta_2 \slashed{p} - (\delta_2 + \delta_m) m_t\big),
\end{align}
with the two-point function before renormalization\footnote{The loop functions $B_0$ and $B_1$ are defined as in Ref.~\cite{Denner:1991kt}.}
\begin{align}
\Sigma^{ab}_t(\slashed{p}) = - \delta_{ab}\frac{m_t^2}{16\pi^2}\frac{\ctt^2}{f_a^2}\left[\slashed{p} B_ 1\left(p^2,m_t^2,m_a^2\right) + m_t B_0\left(p^2,m_t^2,m_a^2\right)\right],
\end{align}
and the renormalization parameters set by ($\Sigma_t^{ab} = \Sigma_t\,\delta_{ab}$)
\begin{align}\label{eq:ren-param}
\delta_m & = \frac{1}{m_t}\Sigma_t(\slashed{p} = m_t) = - \frac{m_t^2}{16\pi^2}\frac{\ctt^2}{f_a^2}\left[B_1(m_t^2,m_t^2,m_a^2) + B_0(m_t^2,m_t^2,m_a^2)\right]\\\nonumber
\delta_1 & = \delta_2 = - \frac{1}{d\slashed{p}}\eval{\Sigma_t(\slashed{p})}_{\mathrm{div}} = - \frac{m_t^2}{32\pi^2}\frac{\ctt^2}{f_a^2}\left(\frac{1}{\epsilon} -\gamma_E + \ln\left(4\pi\right)\right).
\end{align}
Using these results, it is straightforward to cancel the divergences in the amplitudes corresponding to the diagrams in~\figref{fig:diagrams}. Analytic results for the renormalized virtual ALP corrections to $gg\to t\bar{t}$ at $\mathcal{O}(\ctt^2\alpha_s/4\pi)$ are given in App.~\ref{app:loop-results}. We will denote virtual NLO contributions to $t\bar{t}$ observables by $\sigma^{(t\bar{t},1)}$.

An alternative way to renormalize the amplitude is to use operators in the Standard Model Effective Field Theory (SMEFT) as counterterms~\cite{Galda:2021hbr}. In this approach, the UV-divergent ALP contributions act as source terms for the RG evolution of the SMEFT operators. We have checked that the divergences we identify with the renormalization parameters $\delta_m$ and $\delta_2$ agree with the corresponding ALP source terms in Ref.~\cite{Galda:2021hbr}. In top-antitop production, effects of the ALP-top coupling can be renormalized without using counterterms beyond the Standard Model. It is therefore not necessary to work in the SMEFT. In App.~\ref{app:alp-smeft}, we give more details about the SMEFT approach to ALP effects in top-antitop production.

%---------------------------------------------------------------
\paragraph{Real radiation}
The third class of ALP contributions to $gg\to t\bar{t}$ at NLO is the real radiation of an ALP, as shown in the third row of~\figref{fig:diagrams}. Since the radiation of a pseudo-scalar exhibits neither soft nor collinear divergences, these contributions can be treated separately from the virtual corrections. We will denote NLO contributions to observables due to real ALP radiation as $\sigma^{(t\bar{t}a,0)}$. We compute these contributions numerically, as described in Sec.~\ref{sec:observables}.

%---------------------------------------------------------------
\subsection{Observables}
\label{sec:observables}
For general ALP-top and ALP-gluon couplings, the total cross section and any un-normalized kinematic distribution of top-antitop production at the LHC can be written as
\begin{align}\label{eq:cross-section}
\sigma(\mu) = &\ \sigma_{\rm SM} + \frac{\alpha_s^2}{4\pi}\frac{\cgg\ctt}{f_a^2}\,\sigma_{a-\text{SM}}^{(t\bar{t},0)} + \frac{\alpha_s^2}{(4\pi)^2}\frac{\cgg^2\ctt^2}{f_a^4}\,\sigma_{a-a}^{(t\bar{t},0)}\\\nonumber
& + \frac{\alpha_s^2}{4\pi}\frac{\ctt^2}{f_a^2}\,\sigma_{a-\text{SM}}^{(t\bar{t},1)}  + \frac{\alpha_s^2}{4\pi}\frac{\ctt^2}{f_a^2}\,\sigma_{a-a}^{(t\bar{t}a,0)} + \mathcal{O}\left(\frac{\alpha_s^3}{(4\pi)^2}\frac{\cgg\ctt}{f_a^2},\frac{\alpha_s^4}{(4\pi)^3}\frac{\cgg^2}{f_a^2}\right).
\end{align}
Here $\sigma_{\rm SM}$ is the Standard-Model prediction in QCD and $\sigma_{a-\text{SM}}^{(t\bar{t},0)}$ and $\sigma_{a-\text{SM}}^{(t\bar{t},1)}$ denote the interfe\-rence terms of ALP and QCD $t\bar{t}$ amplitudes at LO and NLO in the SM+ALP. Due to the group structure of QCD, the ALP amplitudes only interfere with the SM $t$- and $u$-channel amplitudes.

The contributions $\sigma_{a-a}^{(t\bar{t},0)}$ and $\sigma_{a-a}^{(t\bar{t}a,0)}$ denote ALP-ALP interference terms for $t\bar{t}$ production and for $t\bar{t}a$ production at tree level. They contribute to inclusive $t\bar{t}$ production at NLO in the SM+ALP. We include the ALP-ALP interference $\sigma_{a-a}^{(t\bar{t},0)}$, even though it has a quartic dependence on the ALP couplings. The reason is that this contribution scales as $m_t^2 s/f_a^4$ compared to the SM cross section~\cite{Gavela:2019cmq,Bruggisser:2023npd}. At high partonic center-of-mass ener\-gies $\sqrt{s} \gtrsim f_a$, it can therefore be numerically relevant despite the $1/f_a^4$ scaling. ALP-ALP interference terms not shown in~\eqref{eq:cross-section} either do not feature this energy enhancement or are suppressed by powers of $\alpha_s/4\pi$. We will neglect them in our analysis.

In~\eqref{eq:cross-section}, we have only included NLO contributions of $\mathcal{O}(\ctt^2/f_a^2)$ and neglected NLO contributions scaling with $\cgg\ctt/f_a^2$ or $\cgg^2/f_a^2$. This is a good approximation of the full ALP contributions as long as $|\ctt| > \alpha_s|\cgg|/4\pi$, \emph{i.e.}, in scenarios where the ALP-top coupling is larger than the (loop-induced) ALP-gluon coupling. As we will see in Sec.~\ref{sec:ctt-cgg}, this requirement does not affect the bounds we set on the $\ctt-\cgg$ parameter space.

In Benchmark I, we can obtain the cross section from~\eqref{eq:cross-section} by setting $\cgg \to \ctt/2$. In this case, all ALP contributions scale as $\ctt^2/f_a^2$, except from the ALP-ALP interference at LO, which scales as $\ctt^4/f_a^4$.

In Benchmark II, contributions of $\cgg$ only enter through the RG evolution. Using the leading-log approximation for $\cgg(\mu)$ from~\eqref{eq:ll}, the cross section is approximately given by\footnote{Here we have introduced $\overline{\sigma}$ to distinguish the contributions from $\sigma$ in~\eqref{eq:cross-section}.}
\begin{align}\label{eq:cross-section-benchmark}
\sigma(\mu) = &\ \sigma_{\rm SM} + \frac{\alpha_s^2}{4\pi}\frac{\ctt^2(\Lambda)}{f_a^2}\left(\frac{\alpha_t}{4\pi}\ln\left(\frac{\Lambda^2}{\mu^2}\right)\,\overline{\sigma}_{a-\text{SM}}^{(t\bar{t},0)} + \overline{\sigma}_{a-\text{SM}}^{(t\bar{t},1)} + \overline{\sigma}_{a-a}^{(t\bar{t}a,0)}\right)\\\nonumber
& + \frac{\alpha_s^2}{(4\pi)^2}\frac{\ctt^4(\Lambda)}{f_a^4}\frac{\alpha_t^2}{(4\pi)^2}\ln^2\left(\frac{\Lambda^2}{\mu^2}\right)\,\overline{\sigma}_{a-a}^{(t\bar{t},0)} + \mathcal{O}\left(\frac{\alpha_s^3}{(4\pi)^2}\frac{\alpha_t}{4\pi}\ln\left(\frac{\Lambda^2}{\mu^2}\right)\frac{\ctt^2}{f_a^2}\right).
\end{align}
The LO contribution of ALP-SM interference, $\overline{\sigma}_{a-\text{SM}}^{(t\bar{t},0)}$, is suppressed compared to the NLO contributions. Also, the energy-enhanced ALP-ALP interference term, $\overline{\sigma}_{a-a}^{(t\bar{t},0)}$, is suppressed compared to the other contributions within the energy reach of the LHC. The dominant ALP effects to top-antitop production are therefore due to virtual corrections, $\overline{\sigma}_{a-\text{SM}}^{(t\bar{t},1)}$, and real radiation, $\overline{\sigma}_{a-a}^{(t\bar{t}a,0)}$.

For the SM contribution to the total cross section at 13 TeV center-of-mass energy, we use the NNLO QCD prediction $\sigma_{\rm SM} = 832^{+40}_{-46}\,$pb, obtained using the program {\tt Top++}~\cite{Czakon:2011xx}. For the SM contributions to the differential distributions, we use NNLO QCD predictions from Ref.~\cite{CMS:2021vhb}, referred to as `MATRIX'. We choose these predictions, because our numerical analysis will be mostly based on the measurement of kinematic distributions by CMS, which is published under the same reference.

To obtain predictions for the ALP contributions to top observables, we combine our analytic results for the LO and virtual NLO corrections from Sec.~\ref{sec:lo-nlo} with simulated contributions of real ALP radiation. To compute the latter, we use {\tt MADGRAPH5\_aMC@NLO} v.3.4.2~\cite{Alwall:2014hca} and simulate 600,000 $pp\to t\bar{t}a$ events for several values of $m_a$ and $\ctt/f_a = 1/$TeV at parton level. We then scale these results with $\ctt^2/f_a^2$ and interpolate between scenarios with different ALP masses. In this way, we obtain predictions for arbitrary values of $\ctt/f_a$ and $m_a$.

For all ALP contributions, we use the NNPDF parton distribution functions (PDFs) \linebreak\texttt{NNPDF31\_nnlo\_as\_0118\_notop}~\cite{NNPDF:2017mvq,Buckley:2014ana}, with a corresponding input value of $\alpha_s(m_Z) = 0.118$. This PDF set excludes top data from the PDF fit, which is performed under the Standard Model assumption. The top mass parameter is set to the pole mass $ m_t = 172.5\,$GeV. Our fixed-order calculation leaves us with a remnant dependence on the factorization scale $\mu_F$ and renormalization scale $\mu_R$. We set $\mu = \mu_F = \mu_R$ and choose a dynamical scale
\begin{align}
\mu = \frac12\sum_{f=t,\bar{t}} m_T(f),\qquad m_T^2(f) = m_f^2 + p_T^2(f),
\end{align}
where $f=t,\bar{t}$ are top and antitop quarks at parton level and $m_T(f)$ is their transverse mass. In particular, the ALP couplings $\ctt(\mu)$ and $\cgg(\mu)$, which enter the observables, should be evaluated at this scale. By RG-evolving these couplings up to the cutoff scale using~\eqref{eq:running}, we obtain predictions of observables in terms of ALP couplings $\ctt = \ctt(\Lambda)$ and $\cgg = \cgg(\Lambda)$, as made explicit in~\eqref{eq:cross-section-benchmark}. This allows us to probe the ALP couplings directly at the cutoff scale, where the ALP EFT can be matched onto a more complete theory.

We now present our numerical predictions for the total cross section and differential distributions of top-antitop production in the two benchmark scenarios. We set the ALP mass to $m_a = 10\,$GeV and fix the cutoff scale at $\Lambda = 4\pi$TeV. The impact of the ALP mass and the ALP-gluon coupling on the observables will be addressed in Sec.~\ref{sec:gluon-coupling-and-mass}. The dependence on the cutoff scale $\Lambda$ enters through the RG evolution of the ALP couplings and is moderate, see~\eqref{eq:running-num}.

%%%%%%%%%%%%%%%%%%%%%%%%%%%%%%%%%%%%%%%%%%%%%%%%%%
\paragraph{Total cross section}
\label{sec:total-xs}
For Benchmark I~\eqref{eq:benchmarkI} with $c_{GG}(\Lambda) = 0$, the total cross section for top-antitop production at the LHC with $\sqrt{s} = 13\,$TeV is given by
\begin{align}\label{eq:xs-I}
    \sigma_{t\bar{t}}^{(I)}\ [\text{pb}] & = 832 + \ctt^2(\Lambda)\left(\frac{\rm TeV}{f_a}\right)^2\left(-(0.126)_{a-\text{SM}}^{(t\bar{t},0)} - (0.239)_{a-\text{SM}}^{(t\bar{t},1)} + (0.068)_{a-a}^{(t\bar{t}a,0)}\right).%Cgg_arr = np.linspace(-10,10,num=21)
\end{align}
For Benchmark II~\eqref{eq:benchmarkII} with $\cgg(\Lambda) = 0$, the total cross section reads
\begin{align}\label{eq:xs-II}
    \sigma_{t\bar{t}}^{(I\!I)}\ [\text{pb}] & = 832 + \ctt^2(\Lambda)\left(\frac{\rm TeV}{f_a}\right)^2\left((0.011)_{a-\text{SM}}^{(t\bar{t},0)} - (0.239)_{a-\text{SM}}^{(t\bar{t},1)} + (0.068)_{a-a}^{(t\bar{t}a,0)}\right).
\end{align}
The predictions include all relevant effects described by the cross section formula~\eqref{eq:cross-section}. In both benchmarks, the dominant effect is due to virtual ALP contributions at NLO, $\sigma_{a-\text{SM}}^{(t\bar{t},1)}$. These contributions are negative and deplete the total cross section, regardless of the sign of $\ctt$. The difference between both scenarios lies in the ALP-SM interference term at LO, $\sigma_{a-\text{SM}}^{(t\bar{t},0)}$, which involves the ALP-gluon coupling. In Benchmark I, the LO contributions are comparable to the NLO contributions; in Benchmark II they are RG-induced and sub-leading. The ALP-ALP interference at LO, $\sigma_{a-a}^{(t\bar{t},0)}$, is comparably small and negligible. These observations confirm the qualitative predictions that we made based on~\eqref{eq:cross-section-benchmark}. The dependence on the ALP mass is negligible within the range $0 < m_a \lesssim 200\,$GeV, so that~\eqref{eq:xs-II} and~\eqref{eq:xs-I} apply to all mass scenarios within this range. 

\begin{figure}[t!]
    \centering
    \includegraphics[width=0.49\textwidth]{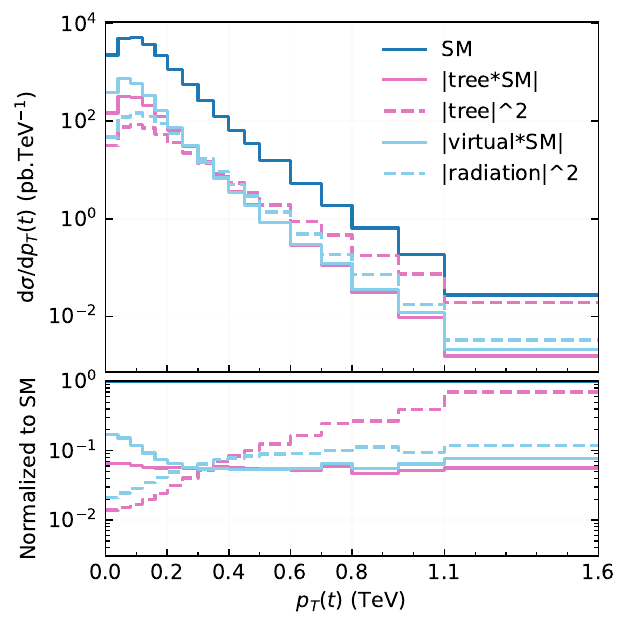}\hfill
    \includegraphics[width=0.49\textwidth]{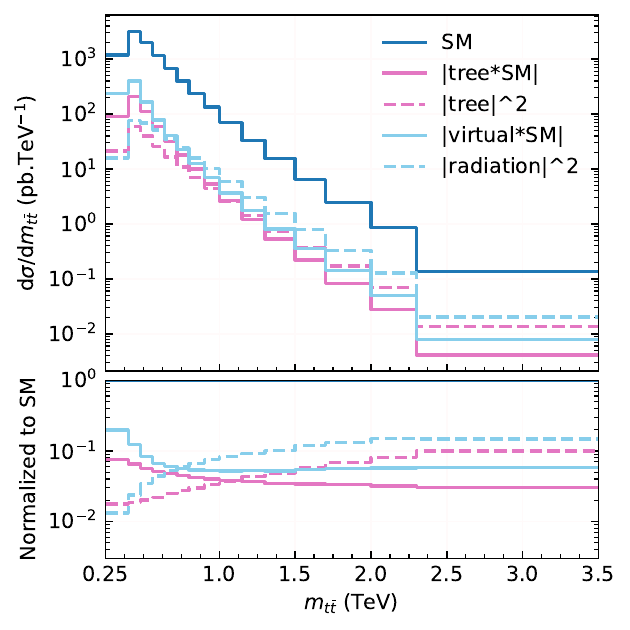}\\[0.3cm]
        \includegraphics[width=0.49\textwidth]{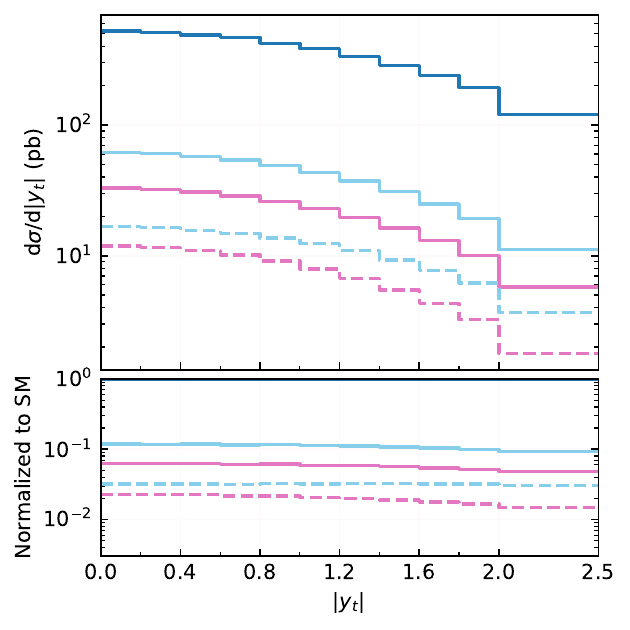}\hfill
    \includegraphics[width=0.49\textwidth]{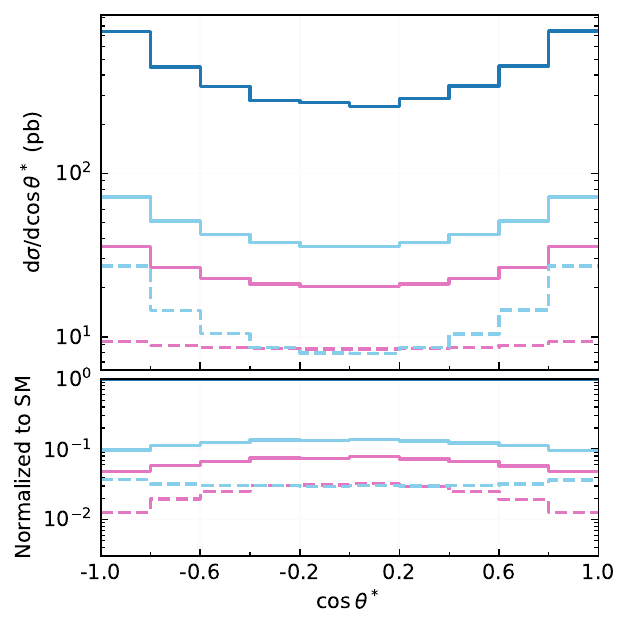}
    \caption{Kinematic distributions in top-antitop production in the SM (blue) and in the SM+ALP for Benchmark I with $c_{tt}/f_a = 20/$TeV, $c_{GG} = 0$ and $m_a = 10\,$GeV. Shown are distributions of the top transverse momentum $p_T(t)$ (upper left), the $t\bar{t}$ invariant mass $m_{t\bar{t}}$ (upper right), the absolute value of the top rapidity $y_t$ (lower left) and the top angle $\theta^\ast$ (lower right). The individual ALP effects correspond to LO contributions from (negative) ALP-SM interference (plain rose) and ALP-ALP interference (dashed rose); as well as NLO contributions from (negative) virtual corrections (plain light blue) and ALP radiation (dashed light blue). \label{fig:distributions-theory}}
\end{figure}

%%%%%%%%%%%%%%%%%%%%%%%%%%%%%%%%%%%%%%%%%%%%%%%%%%
\paragraph{Top-quark kinematic distributions}
\label{sec:distributions}
Besides the total cross section, we study distributions in terms of four kinematic variables: the transverse momentum of the top, $p_T(t)$, the invariant mass of the top-antitop pair, $m_{t \bar t}$, the top-quark rapidity, $y_t$, and the (cosine of the) angle between the top and the flight direction of the $t\bar t$ system in the $t\bar t$ center-of-mass frame, $\cos\theta^\ast$. We focus on Benchmark I, which includes effects from ALP-ALP interference at tree level. Predictions of the four distributions are presented in~\figref{fig:distributions-theory} for $c_{tt}(\Lambda)/f_a = 20/$TeV, $c_{GG}(\Lambda) = 0$ and $m_a = 10\,$GeV.

%%%%%%%%%%%%%%%%%%%%%%%%%%%%%%%%%%%
%%%%%%%%%%%%%%%%%%%%%%%%%%%%%%%%%%%

The various ALP contributions deserve a detailed discussion. In Benchmark I, contributions from ALP-SM interference at LO and NLO are negative in all kinematic distributions.~\footnote{RG-induced contributions from ALP-SM interference at LO are positive, but small, cf.~\eqref{eq:xs-II}.} At energies close to the top-antitop production threshold, virtual NLO corrections dominate the overall ALP effect, as observed in the $p_T(t)$ and $m_{tt}$ distributions in~\figref{fig:distributions}. ALP radiation is suppressed close to threshold, where the phase space to produce extra particles is suppressed. At higher energies, ALP radiation increases and eventually dominates over virtual contributions.

ALP-ALP interference at LO plays a special role for the overall ALP effect. At high energies, it is enhanced over the Standard Model with $m_t^2 s/f_a^4$, as discussed above. In scenarios with non-vanishing ALP-gluon couplings, this enhancement can dominate the overall ALP effect in the tails of the $p_T(t)$ and $m_{tt}$ distributions. In Benchmark I, ALP-ALP interference dominates the $p_T(t)$ distribution at high energies. In the $m_{t\bar{t}}$ distribution, the enhancement of ALP-ALP interference is less pronounced and ALP radiation dominates at high energies.

The shape of the ALP contributions in the $p_T(t)$ and $m_{tt}$ distributions differs significantly from the SM prediction. This indicates that differential distributions provide additional sensitivity to ALPs beyond the total cross section. ALP effects on the rapidity ($y_t$) and angular ($\cos\theta^\ast$) distributions, however, have a very similar shape as the SM prediction and are numerically small, see Fig.~\ref{fig:distributions}.  This implies that the normalized distributions have basically no sensitivity to ALPs. Moreover, the distributions are symmetric under $y_t \leftrightarrow - y_t$ and $\cos\theta^\ast \leftrightarrow - \cos\theta^\ast$, due to the symmetric gluon-gluon initial state. This implies that no top-antitop asymmetry is induced.

In what follows, we will focus our analysis on the $p_T(t)$ and $m_{tt}$ distributions, which promise the highest sensitivity to ALPs. Observables like the transverse momentum of the top-antitop pair might offer additional sensitivity, but are also highly sensitive to QCD radiation. Top polarization observables are interesting, because they can measure spin flips of the top quark, which are induced by the interaction with the ALP. Since polarization observables such as spin correlations are measured through the angular distributions of the top decay products, they require a dedicated analysis based on final-state particles. We leave this interesting direction for future work. 

%%%%%%%%%%%%%%%%%%%%%%%%%%%%%%%%%%%%%%%%%%%%%%%%%%%%%%%%%%%%%%%
\subsection{Impact of the ALP mass and gluon coupling}
\label{sec:gluon-coupling-and-mass}
In this section, we extend our analysis beyond the two benchmarks and study the impact of the ALP mass and the ALP-gluon coupling on the observables. We focus on the $p_T(t)$ distribution, where the effect of the ALP-gluon coupling is most pronounced. The ALP-top coupling is fixed to $\ctt/f_a = 20$/TeV.

\begin{figure}[t!]
    \centering
   \includegraphics[width=0.49\textwidth]{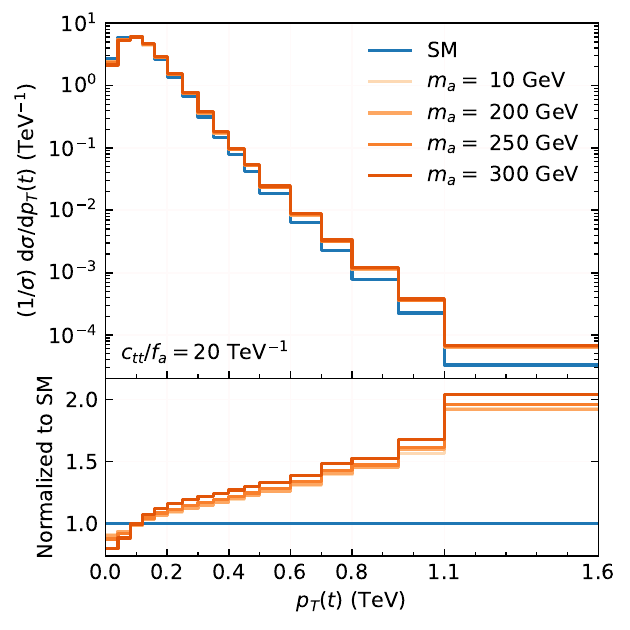}\hfill
 \includegraphics[width=0.49\textwidth]{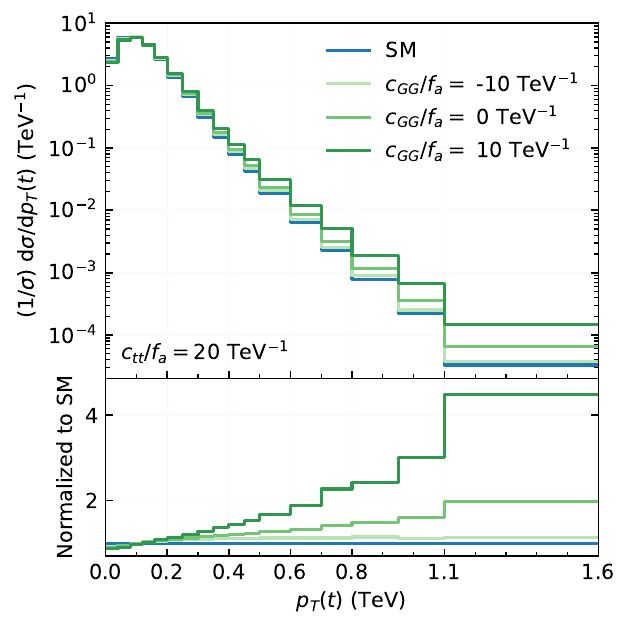}
    \caption{Normalized kinematic distributions of the top transverse momentum for different values of the ALP mass $m_a$ (left) and the ALP-gluon coupling $c_{GG}$ (right). The ALP-top coupling is fixed to $c_{tt}/f_a = 20$ TeV$^{-1}$. In the left panel, the gluon coupling is $c_{GG} = 0$, as in Benchmark I. In the right panel, the ALP mass is fixed to $m_a = 10\,$GeV. \label{fig:ma-variation}}
\end{figure}

In Fig.~\ref{fig:ma-variation}, left, we show the normalized $p_T(t)$ distribution for various fixed ALP masses in Benchmark I. For $0 < m_a \lesssim 200\,$GeV, the distribution is insensitive to the ALP mass. For larger ALP masses, the ALP effect gets stronger as the mass approaches the top-antitop threshold. We do not explore the resonance region, where effects of the ALP width matter and introduce a model dependence. In this region, top-antitop resonance searches might offer further sensitivity.

%%%%%%%%%%%%%%%%%%%%%%%%%%%%%%%%%%%
%%%%%%%%%%%%%%%%%%%%%%%%%%%%%%%%%%%

In Fig.~\ref{fig:ma-variation}, right, we fix the ALP mass to $m_a = 10\,$GeV and vary the ALP-gluon coupling. The variation changes the LO contributions from ALP-SM and ALP-ALP interference compared to the NLO contributions. The values $c_{GG}/f_a = 0$ and $c_{GG}/f_a = -10/$TeV correspond to Benchmarks I and II, respectively. For $c_{GG}/f_a = 10/$TeV, the effective ALP-gluon coupling is $\cgg = c_{GG} + \ctt/2 = 20/$TeV and the tail of the distribution is strongly enhanced by ALP-ALP interference.

Qualitatively, these effects are similar for the $m_{tt}$ distribution.

%%%%%%%%%%%%%%%%%%%%%%%%%%%%%%%%%%%%%%%%%%%%%%%%%%%%%%%%%%
\subsection{Effective coupling approximation}
\label{sec:effective-coupling}
Previous studies of ALP effects in top observables have used approximations of the full NLO contribution~\cite{Galda:2021hbr,Bonilla:2021ufe,Esser:2023fdo,Bruggisser:2023npd}. In particular, studies of the ALP-top coupling have only included the second diagram in the first row of Fig.~\ref{fig:diagrams}. The corresponding contribution to top-antitop production is UV-finite~\cite{Bauer:2020jbp,Bonilla:2021ufe}. At high energies $\sqrt{s} \gg 2 m_t$, the loop function reduces to a constant and the vertex can be treated as an effective ALP-gluon coupling $\cgg = c_{GG} + \ctt/2$. We call the interference between the top-loop diagram from Fig.~\ref{fig:diagrams} and the SM LO amplitude the \emph{effective coupling approximation} of the full ALP effect at NLO.

\begin{figure}[t!]
    \centering
   \includegraphics[width=0.49\textwidth]{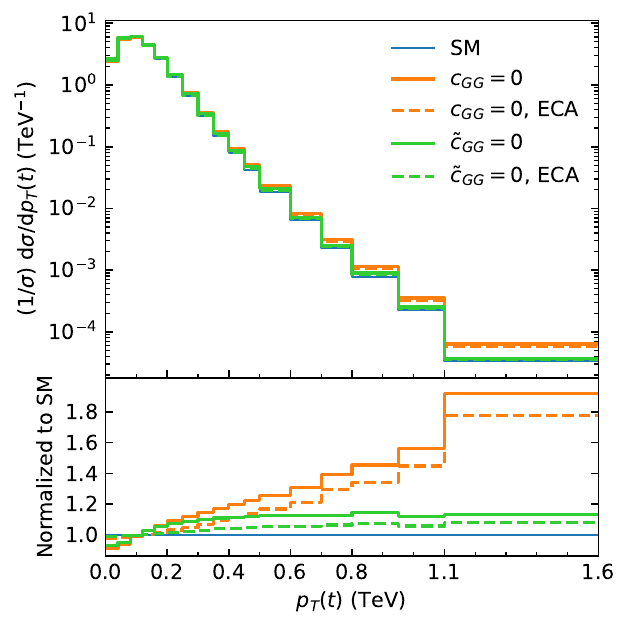}
   \includegraphics[width=0.49\textwidth]{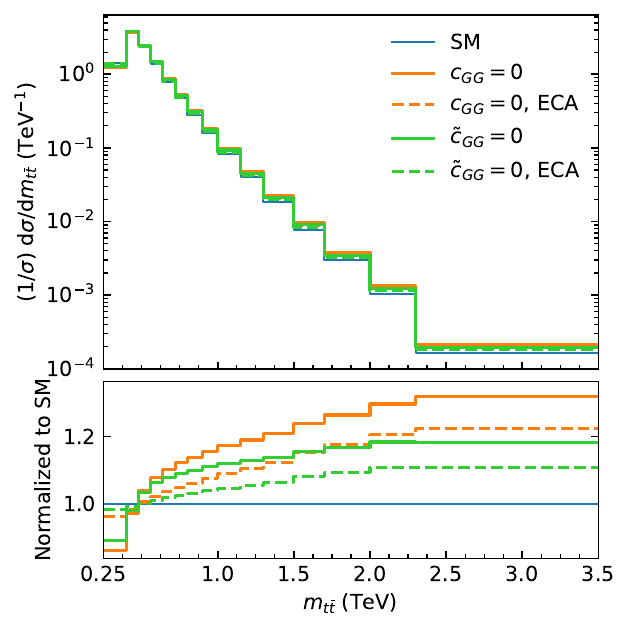}
    \caption{Normalized kinematic distributions of $p_T(t)$ and $m_{tt}$ in the SM+ALP (plain) and in the effective coupling approximation ECA (dashed). The distributions are shown for $\ctt/f_a = 20/$TeV and $m_a = 10\,$GeV in Benchmark I (orange) and Benchmark II (green). \label{fig:eff-coupling}}
\end{figure}

In Fig.~\ref{fig:eff-coupling}, we compare the normalized $p_T(t)$ and $m_{tt}$ distributions in the effective coupling approximation (ECA) with the full NLO result from our calculation. It is apparent that the ECA underestimates the ALP effects in both benchmark scenarios. Including the full virtual corrections and real ALP radiation is numerically important to correctly predict the ALP effects in top-antitop production.

%%%%%%%%%%%%%%%%%%%%%%%%%%%%%%%%%%%
%%%%%%%%%%%%%%%%%%%%%%%%%%%%%%%%%%%

%%%%%%%%%%%%%%%%%%%%%%%%%%%%%%%%%%%%%%%%%%%%%%%%%%%%%%%%%%%%%%%
\section{Comparison with LHC data}
\label{SEC:results}
To compare our predictions with data, we choose to work with measurements of normalized differential distributions in top-antitop production by the CMS collaboration~\cite{CMS:2021vhb}. The analysis is based on $137\,$fb$^{-1}$ of LHC data collected at a center-of-mass (CM) energy of $\sqrt{s} = 13\,$TeV during Run II, using the lepton+jets final state with one leptonically and one hadronically decaying top quark. The tops are reconstructed at parton level, and the analysis provides the relevant information about experimental uncertainties and their correlations between different observables, which is optimal for interpretation. For comparison, we also perform a fit to ATLAS measurements of the same differential distributions~\cite{ATLAS:2019hxz} in the lepton+jets channel, based on $36\,$fb$^{-1}$ of 13-TeV LHC data. Since the statistical combination of multiple measurements relies on experimental uncertainties that are not fully known to us, we refrain from performing a global analysis of LHC data.

\begin{figure}[t!]
    \centering
    \includegraphics[width=0.49\textwidth]{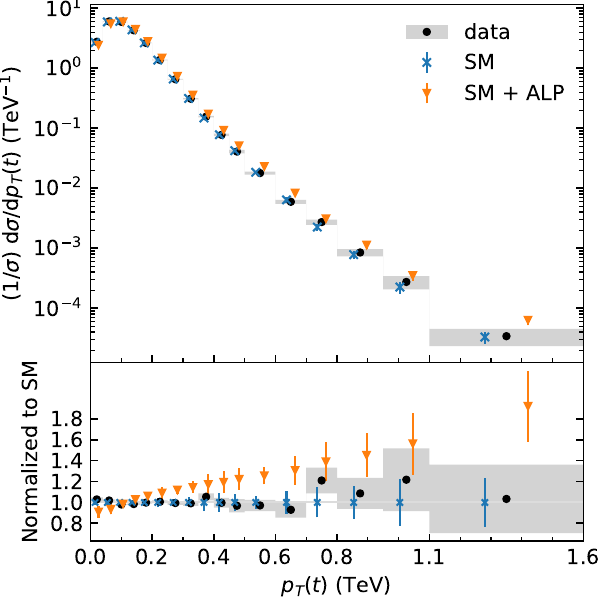}\hfill
    \includegraphics[width=0.49\textwidth]{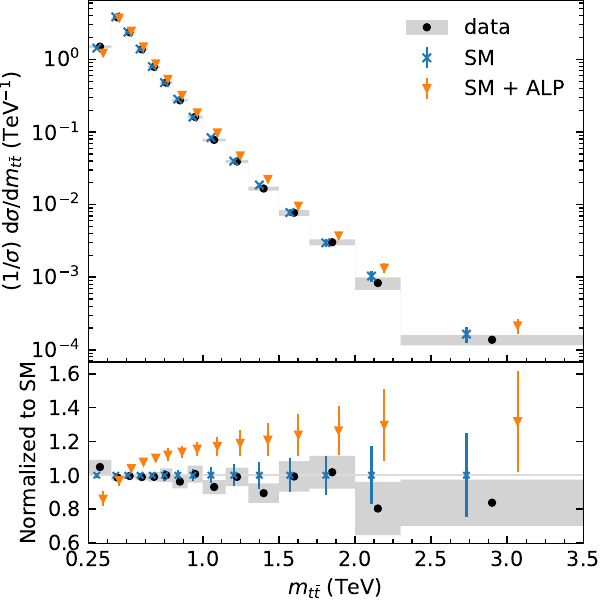}\\[0.4cm]
        \includegraphics[width=0.49\textwidth]{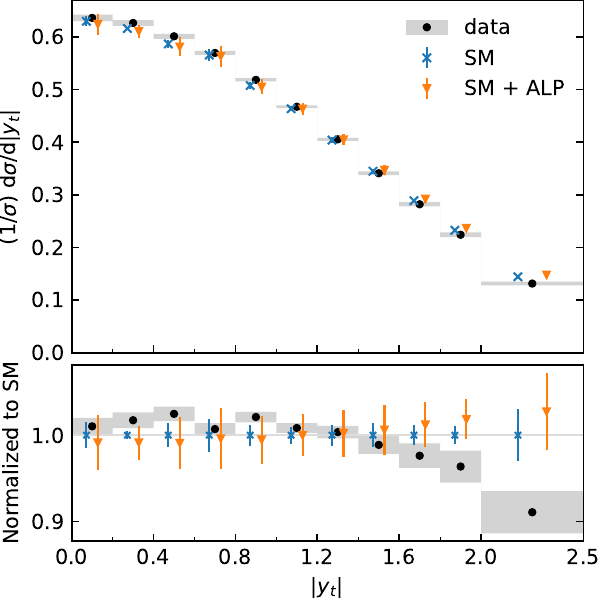}\hfill
    \includegraphics[width=0.49\textwidth]{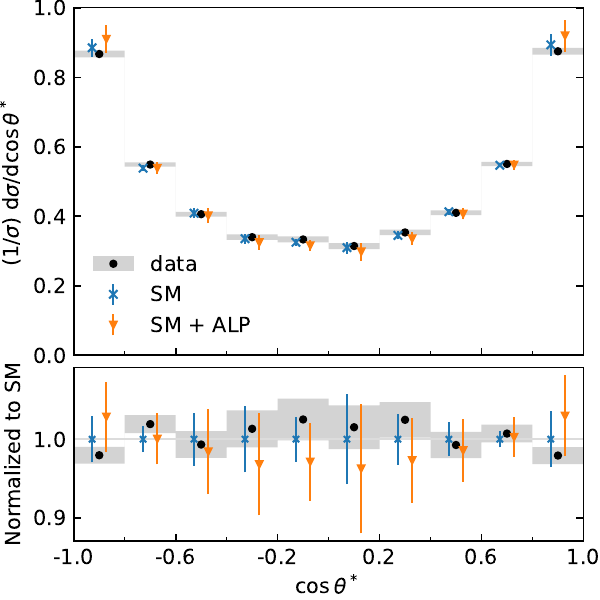}
    \caption{Kinematic distributions in top-antitop production in the SM (blue) and in the SM+ALP (orange) in Benchmark I with $c_{tt}(\Lambda)/f_a = 20/$TeV and $m_a = 10\,$GeV. Show are normalized distributions of the transverse momentum of the (hadronically decaying) top $p_T(t)$ (upper left), the top-antitop invariant mass $m_{t\bar{t}}$ (upper right), the top rapidity $y_t$ (lower left) and the top angle $\theta^\ast$ (lower right). The black dots correspond to CMS measurements~\cite{CMS:2021vhb}; the grey bands indicate the systematic and statistical experimental uncertainties. The blue and orange error bars represent the SM uncertainty $\delta_{\rm SM}$ and the SM+ALP uncertainty $\delta_{\rm th}$, as explained in the text. \label{fig:distributions}}
\end{figure}

In Fig.~\ref{fig:distributions}, we show the four differential distributions discussed in Sec.~\ref{sec:observables} for Benchmark I in comparison with the CMS measurements. The distributions are normalized to the total top-antitop cross section in the SM (blue) and in the SM+ALP (orange). The ALP-top coupling is fixed to $\ctt/f_a = 20/$TeV to illustrate the effects. The ALP contributions are the sum of all individual contributions from Fig.~\ref{fig:distributions-theory}. From the figure, it is apparent that the ALP contributions cause a significant shape variation in the $p_T(t)$ and $m_{tt}$ distributions. The $y_t$ and $\theta^\ast$ distributions, in turn, are essentially insensitive to ALP effects, as discussed in Sec.~\ref{sec:observables}. In our fit to data, we will therefore focus on the $p_T(t)$ and $m_{tt}$ distributions.

%%%%%%%%%%%%%%%%%%%%%%%%%%%%%%%%%%%
%%%%%%%%%%%%%%%%%%%%%%%%%%%%%%%%%%%

In what follows, we perform a fit of our SM+ALP predictions from Sec.~\ref{sec:observables} to data. In Sec.~\ref{sec:fit-setup}, we discuss the setup for our fit and its statistical interpretation. In Sec.~\ref{sec:bounds}, we derive bounds on the ALP-top coupling in Benchmarks I and II, assuming that no other couplings are present at the cutoff scale. In Sec.~\ref{sec:ctt-cgg}, we relax this assumption and derive bounds on the ALP effective theory with arbitrary ALP-gluon and ALP-top couplings.

%%%%%%%%%%%%%%%%%%%%%%%%%%%%%%%%%%%%%%%%%%%%%%%%%%%%%%%
\subsection{Fit setup}
\label{sec:fit-setup}
For the statistical interpretation of our fit, we employ a frequentist approach based on the RFit scheme~\cite{Hocker:2001xe}. This allows us to account for theory uncertainties with a non-Gaussian distribution, while we assume that experimental uncertainties are Gaussian distributed. For $N$ measurements, the log-likelihood function is given by\footnote{For simplicity, we refer to the effective ALP-top coupling as $\ctt$, keeping in mind that the quantity of interest is $\ctt/f_a$.}
\begin{align}
 \chi^2(\ctt,\delta_{\rm th}) = \vec{\chi}^T(\ctt;\delta_{\rm th})\, C^{-1}\, \vec\chi(\ctt;\delta_{\rm th}),
\end{align}
where $C$ is the $N\times N$-dimensional covariance matrix including experimental uncertainties and correlations between individual measurements, in our case bins of distributions. The $i=1,\dots , N$ components of the vectors $\vec{\chi}$ read $\chi_i = |x_i - y_i(\ctt;\delta_{\rm th})|$, where $x_i$ is the central value of measurement $i$ and $y_i(\ctt;\delta_{\rm th})$ is the theory prediction of the corresponding observable in the SM+ALP.

The theory prediction is a function of $\ctt$ and depends on an overall theory uncertainty $\delta_{\rm th}$. For the theory uncertainties of the SM predictions, $\delta_{\rm SM}$, we adopt the values from Ref.~\cite{CMS:2021vhb} in each bin of the normalized kinematic distributions. To take account of the fact that the ALP uncertainties might scale differently, we add an overall uncertainty of $\delta_{\rm ALP}(\ctt) = 0.1\cdot \sigma_{\rm ALP}(\ctt)$, where $\sigma_{\rm ALP}(\ctt)$ stands for the ALP contribution to the observable. The total theory uncertainty is thus given by $\delta_{\rm th} = \delta_{\rm SM} + \delta_{\rm ALP}(\ctt)$.

For fixed values of $\ctt$, we find the minimum log-likelihood function $\chi_{\rm min}^2(\ctt)$ by profiling over the theory predictions within the interval $[\mu_i(\ctt) - \delta_{\rm th}, \mu_i(\ctt) + \delta_{\rm th}]$, where $\mu_i(\ctt)$ is the central value of the theory prediction. Finally, we define our statistical measure as the difference
\begin{align}
\Delta \chi^2(\ctt) = \chi_{\min}^2(\ctt) - \chi_{\rm bf}^2, \qquad \chi_{\rm bf}^2 = \underset{\ctt}{\min}\, \chi_{\mathrm{min}}^2(\ctt),
\end{align}
where $\chi_{\rm bf}^2$ is the best fit with respect to $\ctt$. Using $\Delta \chi^2(\ctt)$ as a measure, rather than $\chi_{\min}^2(\ctt)$, reduces the dependence of the bound on $\ctt$ on a potential mismatch of the SM predictions with the data. All in all, precision measurements of top-quark observables agree well with precision predictions. However, discrepancies can arise from, for instance, mis-modelled signal predictions or underestimated theory uncertainties. Since $\chi_{\min}^2(\ctt)$ follows a Gaussian distribution, we can use $\Delta \chi^2(\ctt)$ directly to derive a bound on $\ctt$ with a certain confidence level (CL). In our results, we quote the 95\% CL bound on $\ctt$, which corresponds to $\Delta \chi^2(\ctt) = 3.84$.

%%%%%%%%%%%%%%%%%%%%%%%%%%%%%%%%%%%%%%%%%%%%%%%%%%%%%%%%%%%%%%%%%%%%%%
\subsection{Bounds on the ALP-top coupling}
\label{sec:bounds}
Using the fitting method described in Sec.~\ref{sec:fit-setup}, we derive bounds on the ALP-top coupling from the top-antitop total cross section and from the differential distributions. The results of our fits are shown in Tab.~\ref{tab:ctt-bounds}.

%%%%%%%%%%%%%%%%%%%%%%%%%%%%%%%%%%%%%%%%%%%%%%%
\begin{table}[tp]\begin{center}
\renewcommand{\arraystretch}{1.2}
\setlength{\tabcolsep}{1.7mm}
\begin{tabular}{l|c|c|c|c|c}
    \toprule
   $|\ctt|/f_a$ (TeV$^{-1}$) & $\sigma_{t\bar{t}}$ & \multicolumn{2}{c|}{$p_T(t)$} & \multicolumn{2}{c}{$m_{tt}$}\\
    & & CMS  &  ATLAS & CMS  &  ATLAS \\
    \midrule
Benchmark I & 16.3 & 14.1 & 17.3 & 11.1 & 7.0 \\
Benchmark II (DFSZ) & 22.5 & 17.7 & 20.8 & 13.0 & 7.1 \\
\bottomrule
\end{tabular}
\end{center}
\caption{95\% C.L. upper bounds on the ALP-top coupling $|\ctt|/f_a$ (TeV$^{-1}$) for ALP masses in the range $0 < m_a \lesssim 200\,$GeV in the SM+ALP. The bounds are obtained from fits of the total top-antitop cross section to four LHC measurements~\cite{CMS:2018fks,CMS:2021vhb,ATLAS:2023gsl,ATLAS:2020aln} and from individual fits of the $p_T(t)$ and $m_{tt}$ distributions to the measurements by CMS~\cite{CMS:2021vhb} and ATLAS~\cite{ATLAS:2019hxz}. \label{tab:ctt-bounds}}
\end{table}
%%%%%%%%%%%%%%%%%%%%%%%%%%%%%%%%%%%%%%%%%%%%%%%

In a first step, we obtain bounds from the total cross section alone. By fitting our predictions from~\eqref{eq:xs-I} and~\eqref{eq:xs-II} to four measurements at the LHC~\cite{CMS:2018fks,CMS:2021vhb,ATLAS:2023gsl,ATLAS:2020aln}, we obtain 95\% C.L. upper bounds on the ALP-top coupling in both benchmark scenarios. To derive bounds from differential distributions, we perform two separate fits of the $p_T(t)$ and $m_{tt}$ distributions to CMS measurements unfolded to parton level~\cite{CMS:2021vhb}. We include all bins shown in Fig.~\ref{fig:distributions}, respecting the bin-to-bin correlations reported in Ref.~\cite{CMS:2021vhb}. We repeat the same procedure for the ATLAS measurements at parton level~\cite{ATLAS:2019hxz}, using the SM predictions quoted in the ATLAS publication.

The invariant mass distribution is most sensitive to ALPs in both benchmark scenarios. Overall, the bounds in Benchmark I are stronger than in Benchmark II. This difference is due to the impact of the effective ALP-gluon coupling, which induces significant tree-level contributions in Benchmark I. Comparing CMS versus ATLAS, we observe that the fit to the $m_{tt}$ distribution from ATLAS sets the strongest bounds on the ALP-top coupling. This result is mostly due to a mismatch between the SM prediction and the measurement. In what follows, we will therefore restrict ourselves to CMS data, in order to obtain conservative bounds.

To investigate the impact of the high-energy tails on the bound, we have repeated the fits to CMS data for Benchmark I without the highest four bins of the distributions. From the $p_T(t)$ and $m_{tt}$ fits, we find $|\ctt|/f_a < 14.1/$TeV and $|\ctt|/f_a < 11.3/$TeV, respectively. This indicates that the bounds on the ALP-top coupling are determined by the low-energy contributions. The energy-enhanced ALP-ALP interference starts dominating only for large ALP-gluon couplings, which are however constrained by dijet distributions, see Sec.~\ref{sec:ctt-cgg}.

Scenarios with $\Lambda = f_a = 1\,$TeV differ from our benchmark by the renormalization group evolution of the ALP couplings from the cutoff scale down to the scales tested in experiments. We have repeated our numerical analysis for $\Lambda = f_a = 1\,$TeV and found that $c_{tt} (m_t)$ is 10\% larger than with $\Lambda = 4\pi\,$TeV. As a consequence, the bound on $\abs{c_{tt}/f_a}$ is strengthened to $7.8 / $TeV, instead of $11.1/$TeV (for light ALPs and $c_{GG} = 0$). However, this result should be taken with care, as for $\Lambda = f_a = 1\,$TeV the cutoff scale lies in the range of the higher bins of the $p_T$ and $m_{t\bar{t}}$ distributions. In this case, the results in the EFT should be interpreted with care.

All these results apply for ALPs with masses in the range $0 < m_a \lesssim 200\,$GeV, where the observables are insensitive to the ALP mass, see Fig.~\ref{fig:ma-variation}. For heavier ALPs, the bounds get stronger as the ALP mass approaches the top-antitop threshold. We will quantify this effect in Sec.~\ref{SEC:comparison}.

%%%%%%%%%%%%%%%%%%%%%%%%%%%%%%%%%%%%%%%%%%%%%%%%%%%%%%%%%%
\subsection{Simultaneous bounds on the ALP-top and ALP-gluon couplings}
\label{sec:ctt-cgg}
Using our predictions from Sec.~\ref{sec:observables} and our study from Sec.~\ref{sec:gluon-coupling-and-mass}, we explore the ALP parameter space in the $\ctt-c_{GG}$ plane. For concreteness, we set the ALP mass to $m_a = 10\,$GeV, keeping in mind that the results will apply to the range $0 < m_a \lesssim 200\,$GeV. As discussed in Sec.~\ref{sec:observables}, our predictions allow us to probe the parameter space for $|\ctt| > \alpha_s|\cgg|/4\pi$. ALP-gluon couplings $|c_{GG}|/f_a \gtrsim 100/$TeV are excluded by angular correlations in dijet production at the LHC~\cite{Bruggisser:2023npd}. This means that our predictions apply as long as $|\ctt|/f_a \gtrsim 1/$TeV. We will see that this is the case in all regions of the parameter space top-antitop production is sensitive to.

To derive bounds on the $\ctt-c_{GG}$ parameter space, we use the total cross section as well as the normalized $p_T(t)$ and $m_{tt}$ distributions. To obtain optimal sensitivity to ALPs throughout the parameter space, we perform two separate fits to data, including the total cross section and either the $p_T(t)$ or the $m_{tt}$ distribution. We do these two separate fits, rather than one combined fit, to avoid double-counting of data in the statistical analysis. We use the same measurements of the total cross section~\cite{CMS:2018fks,CMS:2021vhb,ATLAS:2023gsl,ATLAS:2020aln} and the $p_T(t)$ or $m_{tt}$ distribution from CMS~\cite{CMS:2021vhb} as in Sec.~\ref{sec:bounds}.

\begin{figure}[t!]
    \centering
   \includegraphics[width=0.7\textwidth]{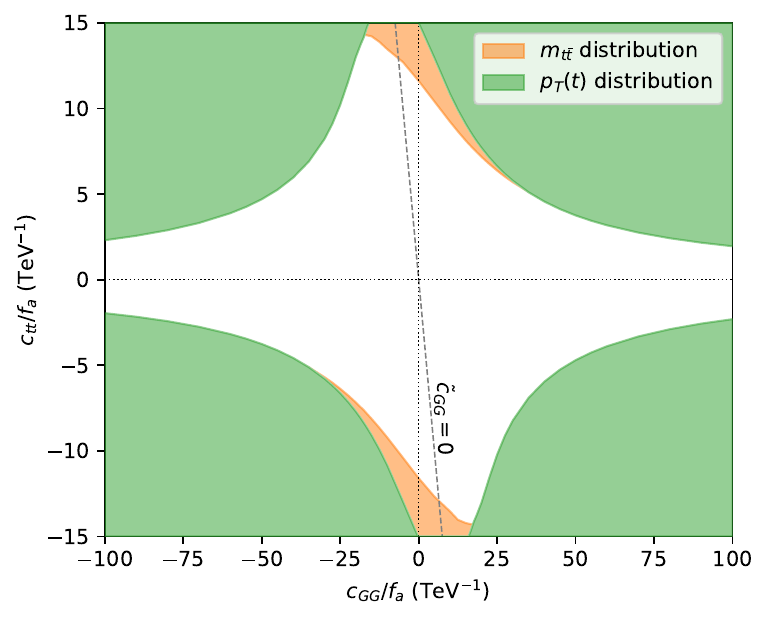}
    \caption{Bounds on the ALP parameter space in the $\ctt - c_{GG}$ plane from top-antitop production. The couplings are defined at the cutoff scale $\Lambda = 4\pi\,$TeV. The colored areas are excluded at 95\% C.L. by measurements of the $p_T(t)$ (green) and $m_{tt}$ (orange) distributions. The bounds apply for ALPs in the mass range $0 < m_a \lesssim 200\,$GeV. \label{fig:ctt-cgg}}
\end{figure}

The results of our fit are shown in Fig.~\ref{fig:ctt-cgg}.
The case $c_{GG} = 0$ corresponds to Benchmark I. For this scenario, the bounds on the ALP-top coupling can be read off from Tab.~\ref{tab:ctt-bounds}. Setting $c_{GG} = -\ctt/2$ or $\cgg = 0$ corresponds to Benchmark II, i.e., to DFSZ-like ALPs. For small ALP-gluon couplings, the $m_{tt}$ distribution dominates the sensitivity. Depending on the sign, the presence of the ALP-gluon coupling at the cutoff scale can either enhance or decrease the sensitivity to the ALP-top coupling. For large ALP-gluon couplings, the $p_T(t)$ distribution is most sensitive, because of the energy-enhanced ALP-ALP interference at LO, see Sec.~\ref{sec:observables}. In this regime, the sensitivity to the ALP-top coupling is significantly enhanced by the presence of the ALP-gluon coupling. This means that in DFSZ-like scenarios with no or a small ALP-gluon coupling at the cutoff scale larger ALP-top couplings are viable compared to scenarios with a large ALP-gluon coupling.

Additional contributions at $\mathcal{O}(f_a^{-4})$ could arise from higher-dimensional operators in the ALP effective theory. They could modify the high-energy tails of distributions, similar to ALP-ALP interference from dimension-5 operators. Once the experimental sensitivity in the tails dominates in the fit, effects of higher-dimensional operators could become visible.

%%%%%%%%%%%%%%%%%%%%%%%%%%%%%%%%%%%
%%%%%%%%%%%%%%%%%%%%%%%%%%%%%%%%%%%

The total cross section is less sensitive to ALP effects than the distributions. In the displayed parameter space, removing it from the fits has no visible effect on the bounds.

%%%%%%%%%%%%%%%%%%%%%%%%%%%%%%%%%%%%%%%%%%%%%%%%%%%%%%%%%%%%%%%
\section{Comparison with other probes of the ALP-top coupling}
\label{SEC:comparison}
In this section, we compare our results from top-antitop production with other collider processes that probe the ALP-top coupling. We consider ALP effects in other top-quark processes at the LHC (Sec.~\ref{sec:other-tops}), searches for light ALPs in rare $B$ meson decays (Sec.~\ref{sec:meson-decays}), and bounds on exotic Higgs decays into ALPs (Sec.~\ref{sec:higgs-decays}). Most of these probes are limited to a certain ALP mass range and/or require additional model assumptions. When comparing their sensitivity with our largely model-independent bounds from top-antitop production, the underlying assumptions should be kept in mind.

%%%%%%%%%%%%%%%%%%%%%%%%%%%%%%%%%%%%%%%%%%%%%%%%%%%%%
\subsection{Other top-quark processes}
\label{sec:other-tops}
Searches for ALP-associated top-antitop production are sensitive to resonant ALPs that decay into various final states. For very light ALPs, the decay happens outside the detector or at a displaced vertex. Heavier ALPs can be detected as promptly decaying resonances.

\paragraph{Top-antitop production plus missing energy}
In Ref.~\cite{Esser:2023fdo}, the authors have re-interpreted a search for $pp\to t\bar{t}\slashed{E}_T$ events with a SUSY stop-neutralino topology by ATLAS~\cite{ATLAS:2021hza}. They find a bound on collider-stable ALPs with top couplings $|\ctt|/f_a < 1.8$/TeV at the 95\% C.L. For ALPs that couple only to top quarks, collider stability can be assumed only for ALP masses well below the di-muon threshold, $m_a < 200\,$MeV, see e.g.~\cite{Rygaard:2023vyo}. In this mass region, missing energy searches can be more sensitive to the ALP-top coupling than inclusive top-antitop production.
 
\paragraph{Top-antitop production with di-muon resonance} Searches for promptly decaying pseudo-scalar resonances have been conducted in $pp \to t \bar{t} a,\, a \to \mu^+\mu^-$ by ATLAS and CMS~\cite{CMS:2019lwf,ATLAS:2023ofo}. The search region targets masses $m_a > 15\,$GeV. For ALPs which couple only to tops, the branching ratio into muon pairs in this mass region is extremely small. The searches are therefore only sensitive to the ALP-top coupling if the ALP branching ratio into muons is sizeable. Alternatively, searches for hadronically decaying pseudo-scalars could probe this mass region up to the $Z$ resonance.

For lighter ALPs that couple only to top quarks, top-antitop production in association with a displaced di-muon vertex is a promising signature. A dedicated analysis predicts that ALPs with masses below the di-bottom threshold can be tested at the LHC~\cite{Rygaard:2023vyo}. With 150/fb of Run-III data, top-associated displaced di-muon searches can probe ALP-top couplings $|\ctt|/f_a < 1/$TeV; with 3/ab at the HL-LHC, the sensitivity increases to $|\ctt|/f_a < 0.1/$TeV.

\paragraph{Four-top production}
A serious competitor to top-antitop production is four-top production at the LHC. ALPs can contribute to this process in multiple ways and through many interfering amplitudes. In Ref.~\cite{Bruggisser:2023npd}, the authors find a bound on the ALP-top coupling, $|\ctt|/f_a \lesssim 4/$TeV, for $m_a = 300\,$MeV from tree-level simulations compared to LHC measurements. Similarly to top-antitop production, the ALP mass dependence in four-top production is supposedly small for ALP masses well below the top-antitop threshold. In Ref.~\cite{Blasi:2023hvb}, the authors find $|\ctt|/f_a \lesssim 6/$TeV under somewhat different assumptions.

These bounds suggests that four-top production is more sensitive to the ALP-top coupling than inclusive top-antitop production. However, making a sound prediction of ALP effects in four-top production is a challenging task, due to the many contributions from individual amplitudes. Moreover, the tree-level analysis of four-top production is subject to rather large theoretical and experimental uncertainties. The resulting bound on the ALP-top coupling should therefore be read with caution. We do not include it in Fig.~\ref{fig:comparison}.

%%%%%%%%%%%%%%%%%%%%%%%%%%%%%%%%%%%%%%%%%%%%%%%%%%%%%
\subsection{Rare $B$ meson decays}
\label{sec:meson-decays}
\begin{figure}[t!]
    \centering
   \includegraphics[width=0.7\textwidth]{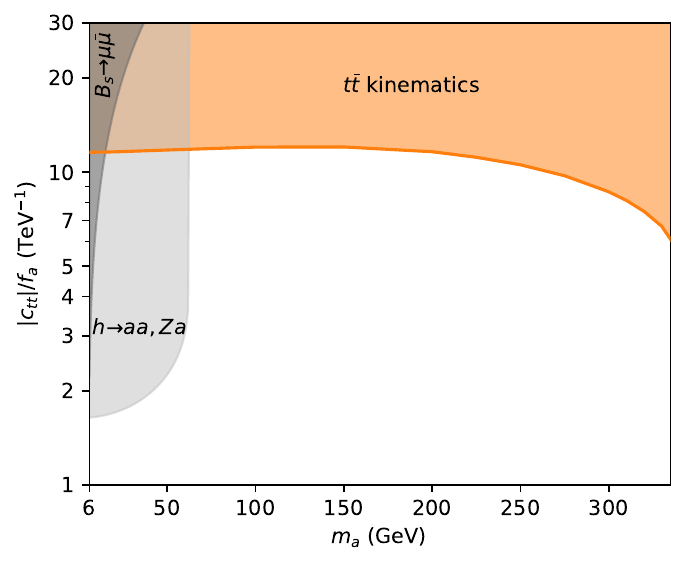}
    \caption{Bounds on the effective ALP-top coupling $|\ctt|/f_a$, defined at the cutoff scale $\Lambda = 4\pi\,$TeV, as a function of the ALP mass $m_a$. Shown are bounds from virtual ALP effects in $B_s\to \mu^+\mu^-$~\cite{Bauer:2021mvw,CMS:2022mgd} (dark grey) and Higgs decays $h\to aa,Za$~\cite{Bauer:2017ris,Bechtle:2014ewa} (light grey), as well as our results from inclusive top-antitop production (orange). The bounds apply to Benchmark I with $c_{GG} = 0$.\label{fig:comparison}}
\end{figure}

ALPs with top couplings can induce rare meson decays through flavor-changing neutral currents at loop level. For ALP masses $m_a < m_B - m_K$, ALPs can be resonantly produced in $B\to K$ decays and searched for in visible and invisible final states. Heavier ALPs can still affect rare meson decays through virtual effects. For $m_a \gtrsim m_B - m_K$, the two-body decay $B_s\to \mu^+\mu^-$ is most sensitive to ALPs with top couplings among the meson decays. The RG evolution and matching of the ALP effective theory generates effective $a\,b\,\bar{s}$ and $a\,\mu\,\bar{\mu}$ couplings at the bottom-mass scale, which induce the decay~\cite{Bauer:2020jbp}. Using the SM predictions for $\mathcal{B}(B_s\to \mu^+\mu^-)$ from Ref.~\cite{Beneke:2019slt} and the ALP contributions from Ref.~\cite{Bauer:2021mvw}, we derive bounds on the ALP-top coupling from LHC measurements. In Fig.~\ref{fig:comparison}, we show the 95\% C.L. upper bound on $\ctt/f_a$ obtained from the most recent measurement by the CMS collaboration~\cite{CMS:2022mgd}. We focus on ALP masses above $6\,$GeV, where resonant production is not possible. For lighter ALPs, resonance searches in flavor processes lead to much stronger bounds, see e.g.~\cite{Bauer:2021mvw,Ferber:2022rsf}.

%%%%%%%%%%%%%%%%%%%%%%%%%%%%%%%%%%%
%%%%%%%%%%%%%%%%%%%%%%%%%%%%%%%%%%%

%%%%%%%%%%%%%%%%%%%%%%%%%%%%%%%%%%%%%%%%%%%%%%%%%%%%%
\subsection{Higgs decays to ALPs}
\label{sec:higgs-decays}
The ALP-top coupling induces the exotic Higgs decays $h\to aZ$ and $h\to aa$ at one-loop level. The corresponding partial decay widths have been calculated in Ref.~\cite{Bauer:2017ris}. The signature of these decay modes strongly depends on the lifetime and mass of the ALP. Dedicated searches for ALP-specific final states offer a high sensitivity, but would also require a separate analysis in each part of the $\{\ctt,m_a\}$ parameter space. Here we derive a conservative bound on the ALP-top coupling, using a bound on un-tagged Higgs decays at the LHC during Run 1, $\mathcal{B}(h \to \text{NP}) < 20\%$ at 95\% C.L.~\cite{Bechtle:2014ewa}. To the best of our knowledge, this is the only interpretation of Higgs measurements that does not rely on further assumptions on the new-physics decay products ``NP''. For ALP masses $m_a < m_h - m_Z$, we interpret this bound as $\mathcal{B}(h \to \text{NP}) = \mathcal{B}(h\to aZ) + \mathcal{B}(h\to aa)$. For the Higgs branching ratios, we assume that at the cutoff scale only the ALP-top coupling is non-zero. In particular, ALP-Higgs and ALP-Higgs-$Z$ couplings that could be induced at higher dimensions in the ALP effective theory are absent. The 95\% C.L. upper bound on the ALP-top coupling is shown in Fig.~\ref{fig:comparison}.\\

From the comparison in Fig.~\ref{fig:comparison}, it is apparent that top-antitop kinematics dominate the sensitivity to the ALP-top coupling for ALPs with masses above $60\,$GeV. Four-top production offers an alternative and similarly sensitive probe. For lighter ALPs, Higgs decays and rare meson decays through virtual tops dominate in sensitivity, but depend heavily on the magnitude of additional ALP couplings. LHC searches for light prompt and displaced resonances in top-antitop production can provide an alternative way to test the ALP-top coupling with on-shell top quarks. They should be conducted for a wide range of masses and in all possible final states.

%%%%%%%%%%%%%%%%%%%%%%%%%%%%%%%%%%%%%%%%%%%%%%%%%%%%%%%%%%%%%%%
\section{Conclusions}
\label{SEC:conclusions}
In this work, we have investigated effects of axions and ALPs in top-antitop production. By calculating all relevant ALP effects at LO and NLO, we have obtained sound predictions of the total cross section and differential distributions. By comparing these predictions to LHC data, we have derived upper bounds on the ALP-top coupling. These bounds apply for a broad range of ALPs with masses below the top-antitop threshold. They constrain scenarios where the ALP-top coupling dominates the phenomenology, such as DFSZ-like axion models.

In a second step, we have extended our analysis to ALPs with arbitrary top and gluon couplings. The resulting bounds on the top-gluon parameter space from top-antitop production constrain the strongly coupling sector of the ALP effective theory. Effects of light quarks are typically suppressed by the small quark masses, due to the intrinsic shift symmetry of the ALP interactions. Therefore the bounds apply to any phenomenologically viable UV completion of the ALP effective theory.

Besides giving valuable information about possible UV completions, bounds on the top and gluon couplings at the cutoff scale limit RG-induced effects on other ALP couplings at lower scales. This is important for the interpretation of low-energy observables, which usually involve all ALP couplings through leading-order or loop-induced effects.

Top-antitop production leads in sensitivity for ALPs with masses from $60\,$GeV up to the top-pair threshold, where searches for top-antitop resonances become available. Lighter ALPs with top couplings have also been probed in Higgs and meson decays, which however rely much more on assumptions on other ALP couplings. Four-top production offers additional sensitivity to ALPs with top couplings across a wide mass range.

For future searches, polarization observables involving the top decay products are promising probes of the ALP-top coupling and definitely worth a study. For light ALPs, searches for displaced vertices in association with top quarks are expected to enhance the sensitivity to the ALP-top coupling compared to inclusive top-antitop production. They can and should be performed at the LHC in all final states of the ALP.

\acknowledgments
We thank Wim Beenakker, Simone Blasi, Sascha Caron, Jochem Kip, Alberto Mariotti, Ken Mimasu, Davide Pagani, Marieke Postma and Simone Tentori for helpful discussions.

%%%%%%%%%%%%%%%%%%%%%%%%%%%%%%%%%%%%%%%%%%%%%%%%%%%%%%%%%%%%%%%
%\clearpage

\appendix
\section{Virtual ALP corrections to $gg\to t\bar{t}$}
\label{app:loop-results}
In this appendix, we give analytic results for the dominant virtual contributions of ALPs to top-antitop production at NLO in the SM+ALP. We include only contributions that are induced by the effective ALP-top coupling $\ctt/f_a$.

We present our results in terms of one-loop amplitudes $\mM^1(gg\to t\bar{t})$. The leading virtual contributions to observables can be obtained by interfering these amplitudes with the tree-level SM amplitudes, which we list here for convenience. For a complete NLO prediction of ALP effects in top-antitop production, these results have to be integrated over the top-antitop phase space and combined with simulations of real ALP radiation, $pp\to t\bar{t} a$.

In what follows, we assign the four-momenta and $SU(3)_C$ indices of the involved particles as
\begin{align}
    g_a(k_1) + g_b(k_2) \to t_i (p_1) + \bar{t}_j (p_2),
\end{align}
where $k_{1,2}$ and $p_{1,2}$ are the four-momenta of the incoming and outgoing particles, and where $a,b,c=\{1,\dots 8\}$ and $i,j,m,n = \{1,\dots 3\}$ are color indices for gluons and quarks.

The gluon polarization vectors and the fermion spinors are abbreviated as
\begin{align}
    \varepsilon_1 \equiv \varepsilon(k_1)
    \ ,\quad
    \varepsilon_2 \equiv \varepsilon(k_2)
    \ ,\quad
    \bar u (p_1) \equiv \bar u_1
    \ ,\quad
    v (p_2) \equiv v_2.
\end{align}
The tree-level amplitudes $\mM^0(gg \to t\bar{t})$ in QCD corresponding to $s$-channel, $t$-channel and $u$-channel Feynman diagrams read
%%%%%%%%%%%%%%%%%%%%%%%%%%%%%%%%%%%%%%%%%%%%%%%
\begin{center}
\begin{tabular}{ | m{2cm}| m{12cm} | }
\hline
    \includegraphics[width=2cm,trim={0.85cm 0 0.95cm 0},clip]{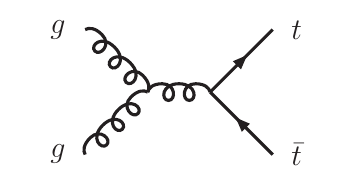}
    & 
    \begin{equation*}
    \begin{aligned}
        \mM^0_s = \frac{g_s^2 i f^{abc} T_{ij}^c}{s}\left[ \left(\varepsilon _1\cdot \varepsilon _2\right) \bar{u}_1\slashed{k}_1v_2 - 2\left(k_1\cdot \varepsilon _2\right) \bar{u}_1\slashed{\varepsilon}_1v_2\right] + \qty(k_1 \leftrightarrow k_2)  \hspace{12cm}
    \end{aligned}
    \end{equation*}\\
  %   \begin{flalign*}
  %       \mM^0_s = \frac{g_s^2 i f^{abc} T_{ij}^c}{s}\left[ \left(\varepsilon _1\cdot \varepsilon _2\right) \bar{u}_1\slashed{k}_1v_2 - 2\left(k_1\cdot \varepsilon _2\right) \bar{u}_1\slashed{\varepsilon}_1v_2\right] + \qty(k_1 \leftrightarrow k_2)
  %       \hspace{1cm}
  % \end{flalign*}\\
\hline
    \includegraphics[width=2cm,trim={0.85cm 0 0.95cm 0},clip]{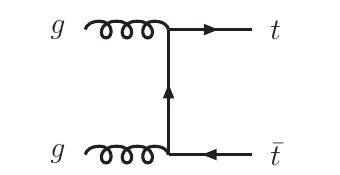}
    & 
    \begin{equation*}
    \begin{aligned}
        \mM^0_t = \frac{g_s^2 T_{im}^a T_{mj}^b}{m_t^2-t} \left[ 2 \left(p_2\cdot \varepsilon _2\right) \bar{u}_1\slashed{\varepsilon}_1v_2 - \bar{u}_1\slashed{\varepsilon}_1\slashed{k}_2\slashed{\varepsilon}_2v_2\right]  \hspace{12cm}
    \end{aligned}
    \end{equation*}\\ 
\hline
    \includegraphics[width=2cm,trim={0.85cm 0 0.95cm 0},clip]{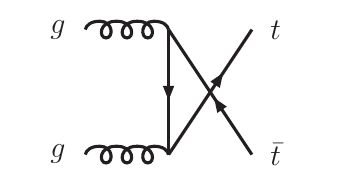}
    &
    \begin{equation*}
    \begin{aligned}
        \mM^0_u = \mM^0_t (k_1 \leftrightarrow k_2)  \hspace{12cm}
    \end{aligned}
    \end{equation*}\\ 
\hline
\end{tabular}
\end{center}
%%%%%%%%%%%%%%%%%%%%%%%%%%%%%%%%%%%%%%%%%%%%%%%%
Here $g_s$ is the strong coupling constant, $T_{ij}^c$ etc. are generators of $SU(3)_C$ and $f^{abc}$ is the structure constant of $SU(3)_C$. We use Einstein's convention to sum over pairs of equal indices. The letters $s,t,u$ are the Mandelstam variables, defined as
\begin{align}
    s & = (k_1 + k_2)^2 = (p_1 + p_2)^2\notag\\
    t & = (k_1 - p_1)^2 = (k_2 - p_2)^2\notag\\
    u & = (k_1 - p_2)^2 = (k_2 - p_1)^2.
\end{align}

The NLO amplitudes for the virtual ALP contributions in the SM+ALP, $\mM^1(gg \to t\bar{t})$, are calculated from the loop diagrams in the first two rows of Fig.~\ref{fig:diagrams}. The numerical contributions of the box diagrams to the top observables considered in this work are about an order of magnitude smaller than the total virtual corrections. We do not include the analytic expressions for them here.

The remaining virtual contributions can be expressed as the sum of re-scaled SM tree-level amplitudes $\mM^0$ and a new structure $\Delta \mM$ that does not occur at tree level. The renormalized amplitudes are given by
%%%%%%%%%%%%%%%%%%%%%%%%%%%%%%%%%%%%%%%%
\begin{center}
\begin{longtable}{ | m{2cm}| m{13cm} | }
\hline
    \includegraphics[width=2cm,trim={0.85cm 0 0.95cm 0},clip]{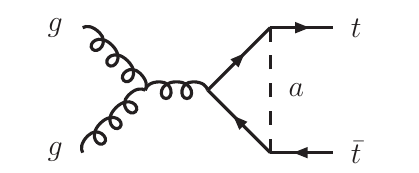}
    & 
    \begin{equation*}
    \begin{aligned}
        \mM^1_s = \frac{c_{tt}^2 m_t^2}{16 \pi ^2 f_a^2}\Big[-c'_{00}\left(s,m_a^2\right) \mM^0_s + c_ {1k}\left(s,m_a^2\right)\qty[\Delta \mM_s(k_1,k_2) + (k_1 \leftrightarrow k_2)]\Big] \label{m1s}
        %\qty[\Delta \mM_s + (k_1 \leftrightarrow k_2)] \label{m1s} 
        \hspace{12cm}
    \end{aligned}
    \end{equation*}
    \\ 
\hline
    \includegraphics[width=2cm,trim={0.85cm 0 0.95cm 0},clip]{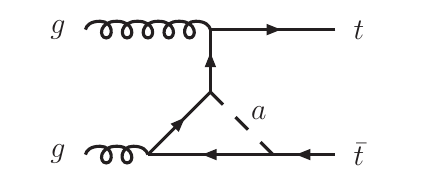}
    &
    \begin{equation*}
    \begin{aligned}
        \mM^1_{t_1} = \frac{c_ {tt}^2 m_t^2}{16 \pi ^2 f_a^2}\Big[- c_ {00}\left(t,m_a^2\right) \mM^0_t +  \Delta \mM_{t}(p_2,k_2)\Big] \hspace{12cm}
    \end{aligned}
    \end{equation*}
    \\ 
\hline
    \includegraphics[width=1.9cm,trim={0.85cm 0 0.95cm 0},clip]{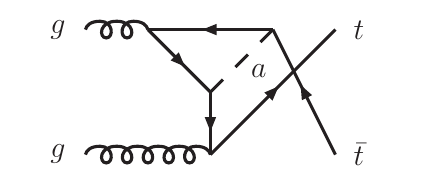}
    & 
    \begin{equation*}
    \begin{aligned}
        \mM^1_{u_1} = \mM^1_{t_1} (k_1 \leftrightarrow k_2) \hspace{12cm}
    \end{aligned}
    \end{equation*}\\
\hline
    \includegraphics[width=1.9cm,trim={0.85cm 0 0.95cm 0},clip]{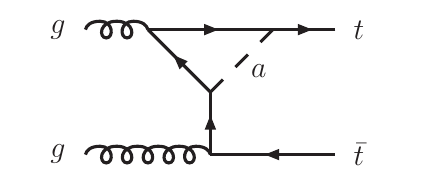}
    &
    \begin{equation*}
    \begin{aligned}
        \mM^1_{t_2} = \mM^1_{t_1}(k_1 \leftrightarrow k_2, p_1 \leftrightarrow p_2)\label{M1t2} \hspace{12cm}
    \end{aligned}
    \end{equation*}
    \\ 
\hline
    \includegraphics[width=1.9cm,trim={0.85cm 0 0.95cm 0},clip]{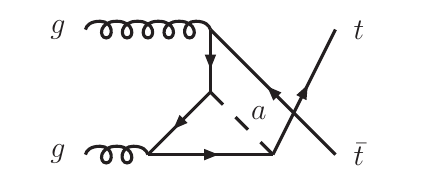}
    &
    \begin{equation*}
    \begin{aligned}
        \mM^1_{u_2} = \mM^1_{t_2}(k_1 \leftrightarrow k_2) = \mM^1_{u_1}(p_1 \leftrightarrow p_2) \hspace{12cm}
    \end{aligned}
    \end{equation*}
    \\ 
\hline
    \includegraphics[width=1.9cm,trim={0.85cm 0 0.95cm 0},clip]{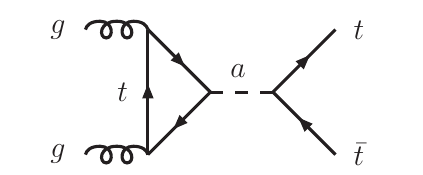}
    &
    \begin{equation*}
    \begin{aligned}
        \mM^1_{a_1} = \frac{g_s^2 c_ {tt}^2  m_t^3}{8 \pi ^2 f_a^2 \left(m_a^2-s \right)}\, \delta_{ij}  \delta^{ab} \,i\,\epsilon^{k_ 1k_ 2\varepsilon_1\varepsilon_2} C_ 0\left(0,s,0,m_t^2,m_t^2,m_t^2\right)  \bar{u}_ 1\bar{\gamma }^5v_ 2 \hspace{12cm}
    \end{aligned}
    \end{equation*}
    \\ 
\hline
    \includegraphics[width=1.9cm,trim={0.85cm 0 0.95cm 0},clip]{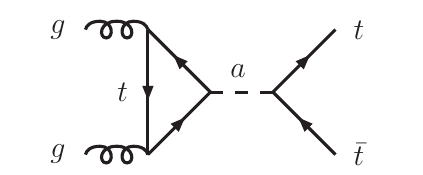}
    &
    \begin{equation*}
    \begin{aligned}
        \mM^1_{a_2} = \mM^1_{a_1} \hspace{12cm}
    \end{aligned}
    \end{equation*}
    \\ 
\hline
    \includegraphics[width=1.9cm,trim={0.85cm 0 0.95cm 0},clip]{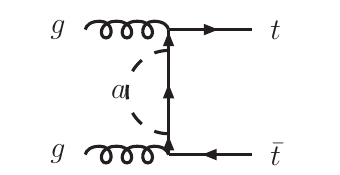}
    &
    \begin{equation*}
    \begin{aligned}
        \mM^1_{t_3} = \frac{c_ {tt}^2 m_t^2}{16 \pi ^2 f_a^2}\Big[ - \frac{m_t^2}{m_t^2-t}b\left(t,m_a^2\right) \mathcal{M}^0_t + g_s^2 T_ {im}^a T_ {mj}^b \,\frac{m_t }{m_t^2-t}\,b'\left(t,m_a^2\right)\, \bar{u}_ 1\slashed{\varepsilon}_ 1\slashed{\varepsilon}_ 2v_ 2 \Big] \hspace{12cm}
    \end{aligned}
    \end{equation*}
    \\ 
\hline
    \includegraphics[width=1.9cm,trim={0.85cm 0 0.95cm 0},clip]{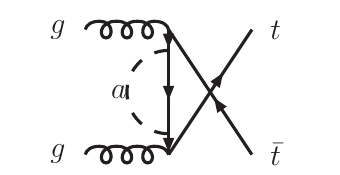}
    &
    \begin{equation*}
    \begin{aligned}
        \mM^1_{u_3} = \mM^1_{t_3}\qty(k_1 \leftrightarrow k_2) \hspace{12cm}
    \end{aligned}
    \end{equation*}
    \\ 
\hline
\end{longtable}
\end{center}
%%%%%%%%%%%%%%%%%%%%%%%%%%%%%%%%%%
 with the new kinematic structures
    \begin{align*}
        \Delta \mM_s (k_1,k_2) &= \frac{g_s^2 i f^{abc} T_ {ij}^c }{m_t s } \left[ 4\left( k_ 1\cdot \varepsilon_2\right) \left(2 m_t \bar{u}_ 1\slashed{\varepsilon}_ 1v_ 2+i \bar{u}_ 1\sigma ^{(k_ 1+k_ 2)\varepsilon_1}v_ 2\right) +  \left(\varepsilon_1\cdot \varepsilon_2\right) (t-u) \bar{u}_ 1v_ 2 \right]\\
        \Delta \mM_{t}(p_2,k_2) &= \frac{g_s^2 T_ {im}^a T_ {mj}^b}{ m_t (m_t^2-t)} \Big[2 \left(p_ 2\cdot \varepsilon_2\right) \left(m_t c_ {0k}\left(t,m_a^2\right) \bar{u}_ 1\slashed{\varepsilon}_ 1v_ 2-c_ {22}\left(t,m_a^2\right) \bar{u}_ 1\slashed{\varepsilon}_ 1\slashed{k}_ 2v_ 2\right)\notag\\
        &\hspace{3cm}+\left(m_t^2-t \right) c_ {2}\left(t,m_a^2\right) \bar{u}_ 1\slashed{\varepsilon}_ 1\slashed{\varepsilon}_ 2v_ 2 \Big].
    \end{align*}
%%%%%%%%%%%%%%%%%%%%%%%%%%%%%%%%%%%%%%%%
Here, the exchange of momenta $k_1\leftrightarrow k_2$ and $p_1 \leftrightarrow p_2$ also exchanges the color indices of the corresponding particles. We have used the shorthand notations
\begin{align}
\sigma ^{(k_ 1+k_ 2)\varepsilon_1} \equiv \sigma_{\mu\nu} (k_1 + k_2)^\mu \varepsilon_1^\nu,\qquad \epsilon^{k_ 1k_ 2\varepsilon_1\varepsilon_2} \equiv \epsilon_{\mu\nu\rho\sigma} k_1^\mu k_2^\nu \varepsilon_1^\mu \varepsilon_2^\nu.
\end{align}
The loop functions are defined as
\begin{align}\label{eq:loop-functions}
    c'_{00}\left(s,m_a^2\right) &= [2 C_{00}\left(m_t^2,s,m_t^2,m_a^2,m_t^2,m_t^2\right) - m_a^2 C_0\left(m_t^2,m_t^2,s,m_t^2,m_a^2,m_t^2\right)\notag\\[0.1cm]
    &\qquad-B_0\left(s,m_t^2,m_t^2\right)]_{\text{fin}}\nonumber\\[0.2cm]
    c_ {1k}\left(s,m_a^2\right) &= m_t^2 \left[C_ {11}\left(m_t^2,s,m_t^2,m_a^2,m_t^2,m_t^2\right)+C_ {12}\left(m_t^2,s,m_t^2,m_a^2,m_t^2,m_t^2\right)\right]\nonumber\\[0.2cm]
    c_ {00}\left(x,m_a^2\right) &= \left[-m_a^2 C_0\left(0,m_t^2,x,m_t^2,m_t^2,m_a^2\right)+2 C_ {00}\left(0,x,m_t^2,m_t^2,m_t^2,m_a^2\right)\right.\notag\\[0.1cm]
    &\quad\left.-B_ 0\left(0,m_t^2,m_t^2\right) + \left(m_t^2-x \right) C_ 1\left(0,x,m_t^2,m_t^2,m_t^2,m_a^2\right)\right]_{\text{fin}} \nonumber\\[0.2cm]
    c_ {2}\left(x,m_a^2\right) &= m_t^2 \left[C_ 0\left(0,m_t^2,x,m_t^2,m_t^2,m_a^2\right)+C_ 2\left(0,x,m_t^2,m_t^2,m_t^2,m_a^2\right)\right]\nonumber\\[0.2cm]
    c_ {22}\left(x,m_a^2\right) &= m_t^2 [C_ 0\left(0,m_t^2,x,m_t^2,m_t^2,m_a^2\right)+2 C_ 2\left(0,x,m_t^2,m_t^2,m_t^2,m_a^2\right) \notag\\[0.1cm]
    &\qquad+C_ {22}\left(0,x,m_t^2,m_t^2,m_t^2,m_a^2\right)]\nonumber\\[0.2cm]
    c_ {0k}\left(x,m_a^2\right) &= -B_1\left(x,m_a^2,m_t^2\right) - m_a^2 C_2\left(0,x,m_t^2,m_t^2,m_t^2,m_a^2\right)+c_ {00}\left(x,m_a^2\right)\nonumber\\[0.2cm]
    b\left(x,m_a^2\right) &= \bigg[\frac{m_t^2-x}{m_t^2} B_ 0\left(x,m_a^2,m_t^2\right)-\frac{m_t^2+x}{m_t^2}  B_ 1\left(x,m_a^2,m_t^2\right) - 2 B_0\left(m_t^2, m_t^2, m_a^2\right) \notag\\[0.1cm]
    &\qquad- 2 B_1 \qty(m_t^2, m_t^2, m_a^2) \bigg]_{\text{fin}}\nonumber\\[0.2cm]
    b'\left(x,m_a^2\right) &= B_1(x,m_a^2,m_t^2) + B_1(m_t^2,m_t^2,m_a^2) + B_0(m_t^2,m_t^2,m_a^2).
\end{align}
The variable $x$ stands for either of the Mandelstam variables $t$ or $u$. The scalar loop functions $B$ and $C$ are defined as in Ref.~\cite{Denner:1991kt}. 

In \eqref{eq:loop-functions}, the label $|_{\text{fin}}$ indicates that terms proportional to $\Delta = 1/\epsilon - \gamma_E + \log 4\pi$ are subtracted from the loop functions due to our renormalization procedure.

%%%%%%%%%%%%%%%%%%%%%%%%%%%%%%%%%%%%%%%%%%%%%%%%%%%%%%%%%%%%%%
\section{Renormalization in the ALP-SMEFT}
\label{app:alp-smeft}
We present an alternative renormalization procedure using SMEFT counterterms, following the strategy proposed in Ref.~\cite{Galda:2021hbr}. In general, this approach renormalizes \emph{all} one-loop amplitudes that are sensitive to the ALP-top coupling $\ctt$. In particular, it can be applied to processes with ALP-top-Higgs couplings, which are not covered by the renormalization procedure in Sec.~\ref{sec:lo-nlo}.

Our goal is to renormalize the one-loop $gg\to t\bar{t}$ amplitudes with ALPs that involve the effective ALP-top coupling. Working in Basis II, the relevant part of the Lagrangian reads (see~\eqref{eq:lagrangianII})
\begin{align}
   \mathcal{L}_{I\!I}(\mu) & \supset - \frac{a}{f_a}\left(\overline{Q}_3 \widetilde{H}\, (\widetilde{\bf{Y}}_U)_{33}\,U_3 + h.c. \right) = - \frac{a}{f_a}\left((\overline{t}_L,\overline{b}_L) \widetilde{H}\,i\, y_t \ctt\,t_R + h.c. \right).
\end{align}
To perform the renormalization, it is convenient to work in a Green's basis. Using the notation from Ref.~\cite{Galda:2021hbr}, the relevant SMEFT Lagrangian in the Green's basis is given by
\begin{align}\label{eq:SMEFT-Lagrange}
    \mL_\SMEFT & \supset \qty(C^{rs}_{uH} Q^{rs}_{uH} + \hc) + \tilde C^{(1)rs}_{Hq} \hat Q^{(1)rs}_{Hq} + \tilde C^{rs}_{Hu} \hat Q^{rs}_{Hu} + \tilde C^{(3)rs}_{Hq} \hat Q^{(3)rs}_{Hq},
\end{align}
where $r,s$ are flavor indices. The SMEFT operators are defined as
\begin{align}\label{eq:operators}
    Q^{rs}_{uH} & = (H^\dagger H)(\overline{Q}_r\widetilde{H}U_s)\\\nonumber
    \hat Q^{rs}_{Hu} & = (H^\dagger H)(\overline{U}_r \stackrel{\longleftrightarrow}{i\slashed{D}} U_s)\\\nonumber
    \hat Q^{(1)rs}_{Hq} & = (H^\dagger H)(\overline{Q}_r \stackrel{\longleftrightarrow}{i\slashed{D}} Q_s)\\\nonumber
    \hat Q^{(3)rs}_{Hq} & = (H^\dagger \sigma^i H)(\overline{Q}_r \stackrel{\longleftrightarrow}{i\slashed{D}}\sigma^i Q_s),
\end{align}
with the SM covariant derivative $D_\mu$ and the Pauli matrices $\sigma^i$.

We use the SMEFT coefficients as counterterms to renormalize the two-point function of the top quark in the SM+ALP. At one-loop order, the renormalized two-point function reads
\begin{align}
    \hat\Sigma_t(\slashed{p}) = \Sigma_t(\slashed{p}) + \frac{v^2}{2}\, \tilde{C}_{V} \slashed p + \frac{v^2}{2}\, \tilde{C}_{A}\, \slashed p \gamma^5 + \frac{v^3}{2\sqrt2}\, C_{uH}^{33},
\end{align}
where $\Sigma_t(\slashed{p})$ is the unrenormalized two-point function and $v = 246\,$GeV is the vacuum expectation value of the Higgs field. The counterterms for the different Lorentz structures are defined as
\begin{align}
\tilde C_V & = \tilde{C}^{(1)33}_{Hq} - \tilde{C}^{(3)33}_{Hq} + \tilde{C}_{Hu}^{33}\\\nonumber
\tilde C_A & = -\tilde{C}^{(1)33}_{Hq} + \tilde{C}^{(3)33}_{Hq} + \tilde{C}_{Hu}^{33}.
\end{align}
They are related to the renormalization parameters $\delta_m$ and $\delta_1$ from~\eqref{eq:ren-param} as 
\begin{align}
    \tilde C_V &= -\frac{2}{v^2}\eval{\dv{\Sigma_t}{\slashed{p}}}_{\mathrm{div}} = \frac{2}{v^2}\,\delta_1 \\\nonumber
    \tilde C_A &= 0\\\nonumber
    C^{33}_{uH} &= - y_t\,\tilde C_V - \frac{2\sqrt 2}{v^3}\,\Sigma_t(\slashed{p} = m_t) = -\frac{2 y_t}{v^2}\,\delta_1 - \frac{2\sqrt 2\,m_t}{v^3}\,\delta_m.
\end{align}
Due to the Ward identity $\delta_1 = \delta_2$, $\tilde C_V$ also renormalizes the top-antitop-gluon vertex.

Instead of the Green's basis in~\eqref{eq:SMEFT-Lagrange}, one can also perform the renormalization with SMEFT operators in the Warsaw basis. In this case, the operator $Q_{uH}^{rs}$ from~\eqref{eq:operators} is the only possible counterterm to renormalize the $gg\to t\bar{t}$ amplitude. In particular, the counterterm $\tilde C_V$ to renormalize the $\slashed p$-dependent part of $\Sigma_t(\slashed p)$, as well as the vertex corrections, is absent. The corresponding divergences are absorbed into the top field strength~\cite{Passarino:2016saj,Kallosh:1973}.
We have explicitly checked that the top field-strength counterterm $\delta_1^{W}$ and the Wilson coefficient $C_{uH}^{33,W}$ in the Warsaw basis are enough to cancel the UV divergences in $gg\to t\bar{t}$, with
\begin{align}
    C_{uH}^{33,W} &= 
    %C^{33}_{uH} + y_t \tilde C_V =
    - \frac{2\sqrt 2\,m_t}{v^3}\,\delta_m, \qquad
    \delta_1^{W} = \delta_1.
\end{align}

%%%%%%%%%%%%%%%%
%%%%%%%%%%%%%%%%
\bibliographystyle{JHEP_improved}
\bibliography{main}
%%%%%%%%%%%%%%%%
%%%%%%%%%%%%%%%%

\end{document}